\documentclass[twocolumn,amsmath,amssymb,prl,superscriptaddress,longbibliography]{revtex4-1}
\usepackage{longtable}
\usepackage{bm}
\usepackage{dcolumn}
\usepackage{epsfig}
\usepackage{graphicx}
\usepackage{subfigure}
\usepackage{comment}
\usepackage{multirow}
\usepackage{float}
\usepackage{calc}
\usepackage{epstopdf}
\usepackage{amsfonts}
\usepackage{amsmath}
\usepackage{gensymb}
\usepackage{amssymb}
\usepackage{subfigure}
\usepackage{color}
\usepackage{epstopdf}
\usepackage{csquotes}
\usepackage{romannum}

\usepackage[colorlinks,bookmarks=false,citecolor=blue,linkcolor=blue,urlcolor=blue]{hyperref}

\newcommand{\be}{\begin{equation}}
\newcommand{\ee}{\end{equation}}
\newcommand{\bea}{\begin{eqnarray}}
\newcommand{\eea}{\end{eqnarray}}

\usepackage{times}
\usepackage{appendix}
\pdfoutput=1

\begin{document}

\title{Superexchange interactions between spin-orbit-coupled $j\!\approx\!1/2$ ions\\
       in oxides with face-sharing ligand octahedra}

\author{Lei Xu}
\affiliation {Institute for Theoretical Solid State Physics, IFW Dresden, Helmholtzstr.~20, 01069 Dresden, Germany}

\author{Ravi Yadav}
\affiliation {Institute for Theoretical Solid State Physics, IFW Dresden, Helmholtzstr.~20, 01069 Dresden, Germany}

\author{Viktor Yushankhai}
\affiliation{Joint Institute for Nuclear Research, Joliot-Curie 6, 141980 Dubna, Russia}
\affiliation{Max-Planck-Institut f\"ur Physik komplexer Systeme, N\"othnitzerstr.~38, 01187 Dresden, Germany}

\author{Liudmila Siurakshina}
\affiliation{Joint Institute for Nuclear Research, Joliot-Curie 6, 141980 Dubna, Russia}

\author{Jeroen van den Brink}
\affiliation {Institute for Theoretical Solid State Physics, IFW Dresden, Helmholtzstr.~20, 01069 Dresden, Germany}
\affiliation{Department of Physics, Technical University Dresden, Helmholtzstr.~10, 01069 Dresden, Germany}
\affiliation{Department of Physics, Washington University, St. Louis, Missouri 63130, USA}

\author{Liviu Hozoi}
\affiliation {Institute for Theoretical Solid State Physics, IFW Dresden, Helmholtzstr.~20, 01069 Dresden, Germany}

\begin{abstract}
Using {\it ab initio} wave-function-based calculations, we provide valuable insights with regard
to the magnetic exchange in 5$d$ and 4$d$ oxides with face-sharing ligand octahedra,
BaIrO$_3$ and BaRhO$_3$.
Surprisingly strong antiferromagnetic Heisenberg interactions as large as 400 meV are computed for
idealized iridate structures with 90$^{\circ}$ Ir-O-Ir bond angles and in the range of 125 meV
for angles of 80$^{\circ}$ as measured experimentally in BaIrO$_3$.
These estimates exceed the values derived so far for corner-sharing and edge-sharing systems
and motivate more detailed experimental investigations of quantum magnets with extended 5$d$/4$d$
orbitals and networks of face-sharing ligand cages.
The strong electron-lattice couplings evidenced by our calculations suggest rich phase diagrams as
function of strain and pressure, a research direction with much potential for materials of this type.

\end{abstract}

\date\today
\maketitle
\section{Introduction}
The interest in the preparation and characterization of 5$d$ oxides and halides goes back to
the 1950's but some of the major implications of having a strong spin-orbit coupling (SOC), at least
for certain 5$d^n$ electron configurations, have been only recently realized.
The work of Kim {\it et al.} on the square-lattice 5$d^5$ iridate Sr$_2$IrO$_4$ \cite{IrO_mott_kim_08,IrO_mott_arima_09},
for example, led to the concept of a spin-orbit driven (Mott-like) insulator while Jackeli and
Khaliullin \cite{Ir213_KH_jackeli_09} brought to the forefront of oxide research the honeycomb
5$d^5$ iridates, as possible hosts for Kitaev physics \cite{Kitaev_2006} and novel magnetic ground states
and excitations \cite{Jeroen_RevModPhys2015}.
Both types of these iridate structures -- square and honeycomb lattices -- have been the topic of
extensive investigations in recent years.
The honeycomb compounds display edge-sharing ligand octahedra and advanced electronic-structure calculations
indicate that the Kitaev exchange is indeed the largest intersite magnetic coupling
\cite{Ir213_yamaji_2014,Na2IrO3_vmk_14}.
Remarkably large anisotropic interactions were also found for corner-sharing ligand cages in
Sr$_2$IrO$_4$, in that case of Dzyaloshinskii-Moriya type, with strengths in the range of 10--15 meV \cite{Ir213_KH_jackeli_09,Ir214_bogdanov_15}.

In contrast to the cases of corner- and edge-sharing coordination, little is known with respect to
the magnitude of the effective coupling constants for adjacent octahedra connected through a O$_3$
facet.
Representative materials of the latter type are the canted antiferromagnet BaIrO$_3$
\cite{Lindsay_BaIrO3_1993,Cao_BaIrO3_2000,Brooks_BaIrO3_2005,Nakano_BaIrO3_2006,Cheng_BaIrO3_2009,Laguna_BaIrO3_2010},
the putative spin-liquid Ba$_{3}$InIr$_{2}$O$_{9}$ \cite{Dey_2017},
the spin-gapped system Ba$_3$BiIr$_2$O$_9$ \cite{Miiller_Ir2O9_2012}, BaRhO$_3$ \cite{Stitzer_BaRhO3_2004},
and BaCoO$_3$ \cite{Sugiyama_Co_2006}.
Here we provide {\it ab initio} results with regard to the strength of facet-mediated superexchange
for IrO$_6$ (RhO$_{6}$) octahedra as found in the 5$d$ (4$d$) $t_{2g}^5$ system BaIrO$_3$ (BaRhO$_{3}$).
We predict remarkably large antiferromagnetic (AFM) Heisenberg interactions in the range of 100 meV
for Ir-O-Ir angles of about 80$^{\circ}$ as found experimentally in BaIrO$_3$ \cite{Stitzer_BaRhO3_2004}.
Moreover, for bond angles $\gtrsim$85$^{\circ}$ the Heisenberg $J$ even exceeds 200 meV in our simulations.
So strong AFM superexchange has been found so far only in one-dimensional corner-sharing cuprates \cite{Suzuura_1996,Motoyama_1996}.
Our findings point to a picture of unusually large, AFM couplings within the face-sharing octahedral units of BaIrO$_3$.
The strong dependence on bond angles of the effective magnetic interactions further resonates with
available experimental data on BaIrO$_3$ \cite{Cao_2004,Cao_BaIrO3_2000,Cheng_BaIrO3_2009,Korneta_2010,Laguna_2014} and Ba$_3$BiIr$_2$O$_3$ \cite{Miiller_Ir2O9_2012},
that indicate subtle interplay between the electronic and lattice degrees of freedom.
\vspace{-0.4cm}
\section{Material model}
\begin{figure*}[t]
 \centering
 \vspace{-0.5cm}
 \subfigure{
\raisebox{7.2cm}{\textbf{ \large (a)}} %\hspace{0.4cm}
 \includegraphics[width=0.45\textwidth]{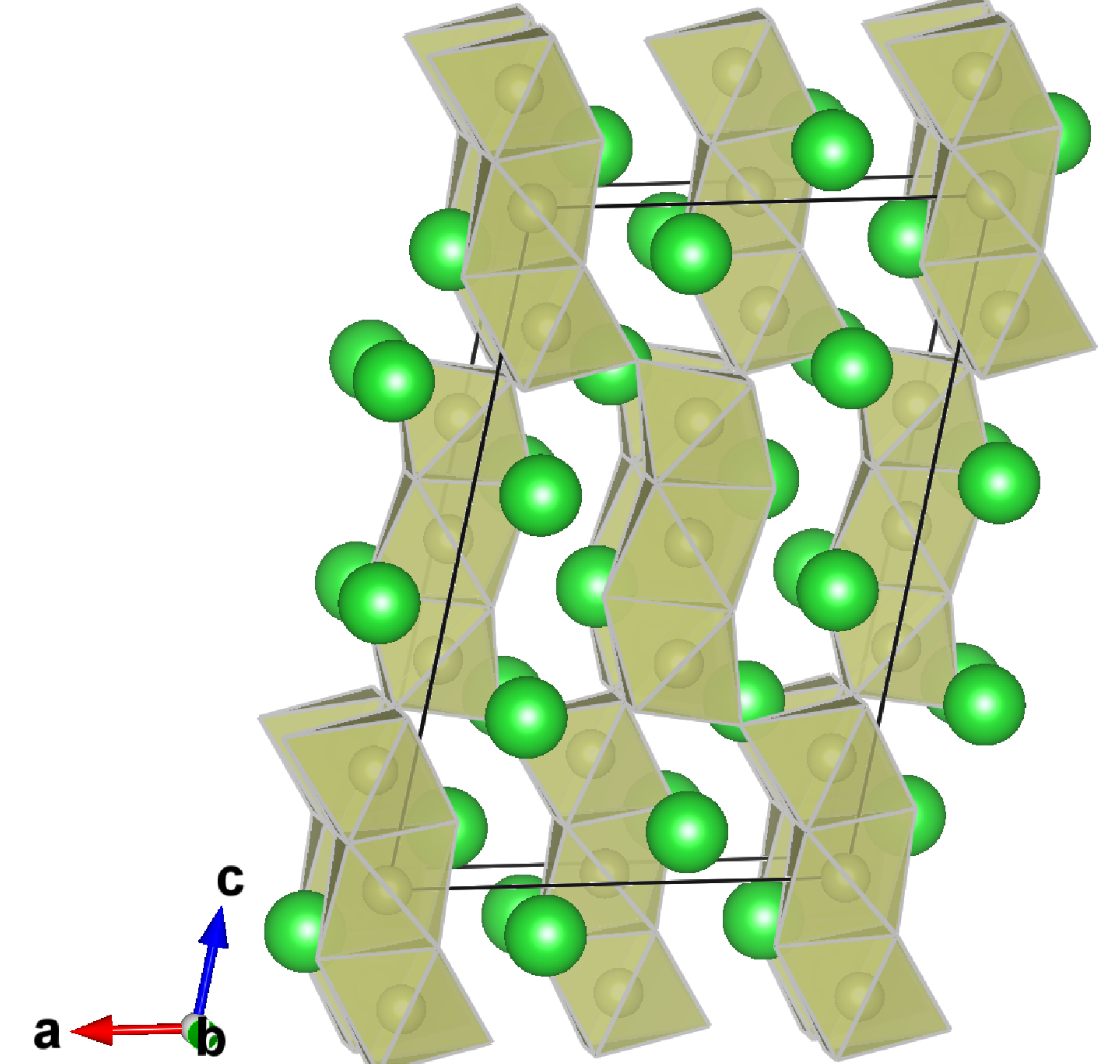}
\label{crystal_BaIrO3}}
 \quad
\subfigure{
\raisebox{7.2cm}{\textbf{ \large (b)}}
 \includegraphics[width=0.38\textwidth,scale=0.2]{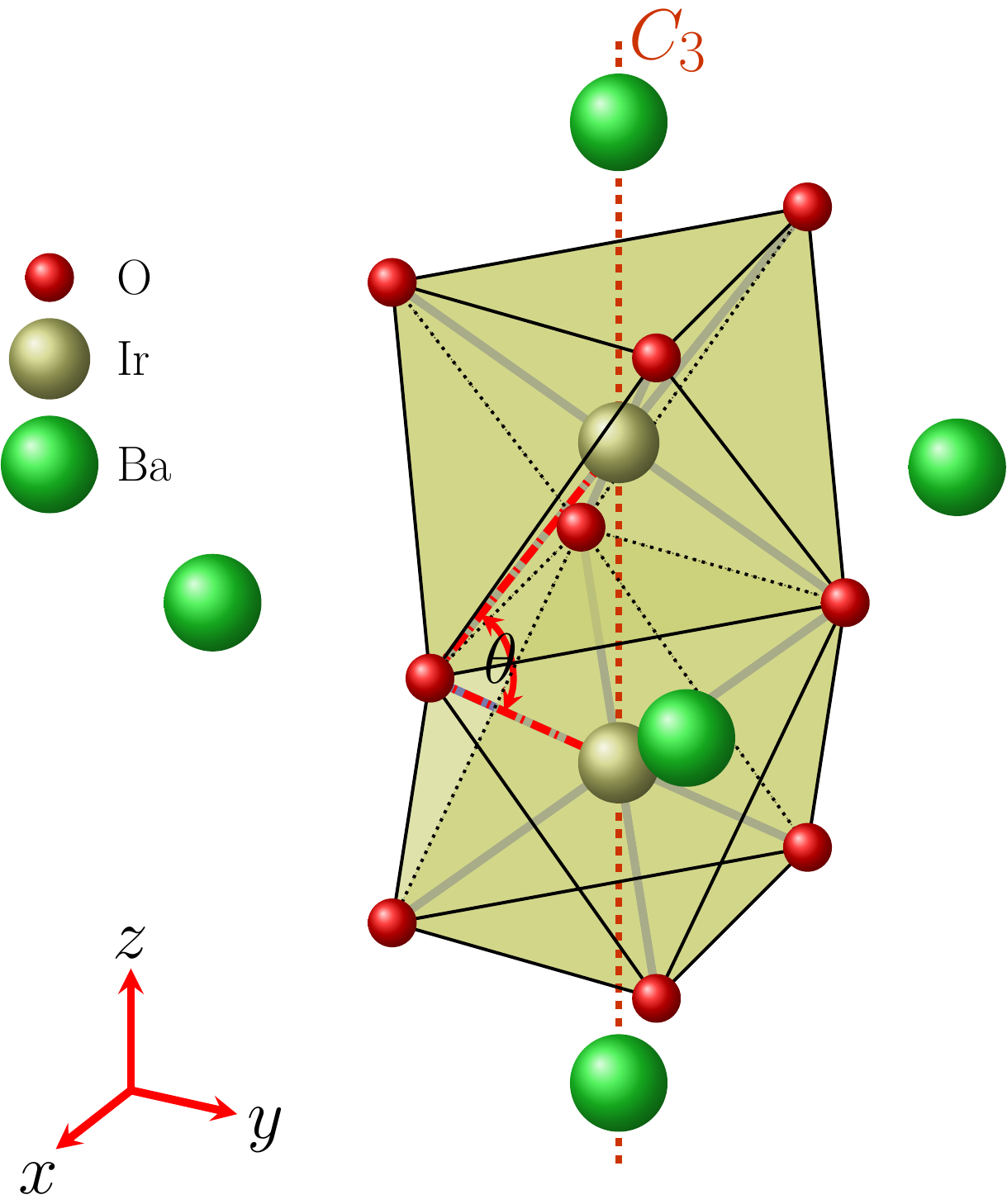}
\label{material_model}} \\
 \caption{(a) Crystal structure of BaIrO$_{3}$, with both face-sharing and corner-sharing IrO$_{6}$ octahedra. Ba atoms are shown as large green spheres.
 (b) Material model used for the calculation of magnetic interactions between two NN TM sites with face-sharing connectivity of the O octahedra.
The point-group symmetry is $D_{3h}$.
   }
\end{figure*}

BaIrO$_3$ features a distorted hexagonal structure with both face-sharing and corner-sharing
 IrO$_{6}$ octahedra \cite{Yuan_BaIrO3_2016}. Those connected by one single ligand form honeycomb-like planes; the linkage
 of adjacent honeycomb layers is ensured by inter-layer Ir ions located such that blocks of three face-sharing
 octahedra are formed along the $c$ axis, see Fig.\,\ref{crystal_BaIrO3}. Since for any pair of nearest-neighbor (NN) octahedra the actual
 point-group symmetry is very low, we focus in our study on an idealized material model  displaying $D_{3h}$ symmetry:
 [Ir$_{2}$O$_{9}$]$^{10-}$ units as depicted in Fig.\,\ref{material_model} around which we additionally considered,
 for keeping overall charge neutrality, three Ba sites within the plane of the median O$_{3}$
 facet and two extra Ba ions along the $z$ axis.
 Although this material model is somewhat oversimplified, it should rather well describe the essential short-range
 electron interactions, as confirmed by similar investigations of edge-sharing $5d^{5}$ compounds \cite{Na2IrO3_vmk_14}.

One feature of 5$d$ transition-metal (TM) ions is that their valence orbitals are much more diffuse
 as compared to first-series TM species. The ligand field is therefore more effectively felt and for instance the Ir$^{4+}$
 ions tend to adopt low-spin $t^{5}_{2g}$ configurations. The more extended nature of the 5$d$ functions further gives rise to large intersite hoppings and
 large superexchange, as in e.g. Sr$_{2}$IrO$_4$ \cite{Ir213_KH_jackeli_09,Ir214_bogdanov_15} and CaIrO$_3$ \cite{Nikolay_CaIrO3_2012}.
 %Another feature of the $5d$ iridates is that the strong relativistic spin-orbit coupling (SOC) entangles large superexchange,
 %which induces large anisotropic terms.
Under strong octahedral crystal fields (CFs) and spin-orbit interactions, with one single unpaired electron ($S$=1/2) in the $t_{2g}$ manifold (orbital angular momentum $L$=1),
the 5$d^{5}$ (4$d^{5}$) valence electron configuration of Ir$^{4+}$ (Rh$^{4+}$)
in BaIrO$_3$ (BaRhO$_3$) yields  an effective $j$=1/2 Kramers-doublet ground state \cite{Ir213_KH_jackeli_09, book_abragam_bleaney}.
Deviations from a perfect cubic environment may lead to some degree of admixture between the $j$=1/2 and lower-lying $j$=3/2 spin-orbit states \cite{book_abragam_bleaney}.
In BaIrO$_{3}$ and BaRhO$_{3}$, in particular, the trigonal distortion of the oxygen octahehra plays a quite important role in this regard, as illustrated in Appendix A
through simple analytical expressions based on an effective ionic model.
To estimate the strengths of the exchange interactions in BaIrO$_3$ and BaRhO$_3$, both isotropic and anisotropic, we here
employ many-body {\it ab initio} techniques from wave-function-based quantum chemistry (QC), then map the magnetic spectrum obtained in the
QC calculations onto an appropriate effective spin Hamiltonian, the form of the latter being
dictated by the symmetry of the material model.
%
%The more extended nature of the 5$d$ functions further gives rise to large intersite hole  hoppings, including both the direct $d-d$ and indirect (superexchange) ones,
 %whose values are very sensitive to the local geometry of the face-sharing bond between NN metal ions.
 %Our QC calculations are also aimed at establishing relations between the exchange coupling and hopping integrals.  The expected relations
 %between these two quantities are discussed in some details in Appendix A where the underlying electronic model is present.
%Our study is also aimed at explaining the trends found numerically in the QC frame on the basis of a simplified superexchange effective
%model. The latter is detailed in Appendix A.

%%%%%%%%%%%%%%
%% FIGURE 1 %%
%%%%%%%%%%%%%%

%Under strong octahedra CF and SOC interactions, with single unpaired electron in the $t_{2g}$ manifold (orbital angular momentum $L$=1),
%the 5$d^{5}$ (4$d^{5}$) valence electron configuration of Ir$^{4+}$ (Rh$^{4+}$)
%in BaIrO$_3$ (BaRhO$_3$) yields  an effective $j\!\approx\!$ 1/2 Kramers-doublet ground state \cite{Ir213_KH_jackeli_09}.
%
\section{Magnetic interactions}
For the idealized M$_{2}$O$_{9}$ cluster (M=Ir, Rh) of face-sharing  octahedra (Fig.\,\ref{material_model})
the overall symmetry is $D_{3h}$.
Each particular superexchange path M$_i$-O$_n$-M$_j$ ($n$ \!=\! $1,2,3$) implies a finite
Dzyaloshinskii-Moriya (DM) vector $\mathbf{D^{\it{n}}_{\it{ij}}}$, since there is no inversion center for the M$_{2}$O$_{9}$ unit.
However, given the $D_{3h}$ symmetry, these DM vectors lie within the plane of the O$_{3}$-facet and are
related to each other through rotations around the $C_{3}$ axis. This yields a vanishing DM coupling
$\mathbf{D_{\it{ij}}}$ \!=\! $\sum_n \mathbf{D^{\it{n}}_{\it{ij}}}$ \!=\! 0.
%
%$\mathbf{D_{\it{ij}}}(n)$ should be permitted. Because of $D_{3h}$ symmetry, these DM vectors are lying in the O$_3$-facet plane and converting  to each other under the rotations around the $C_3$ axis, which results
% in zero total  $\mathbf{D_{\it{ij}}}$ \!=\! $\sum_n \mathbf{D_{\it{ij}}}(n)$ \!=\! 0.
%thus the DM anisotropy term vanishes.
For a pair of NN 1/2 pseudospins
$\bf{\tilde{S}_{\it{i}}}$ and $\bf{\tilde{S}_{\it{j}}}$ with this type of linkage, the most general bilinear spin Hamiltonian can be then cast in the form
\begin{equation}
 {\cal H}_{ij} =   \mathbf{{\it J}}_{ij} \mathbf{\tilde{S}_{\it{i}}}\cdot \mathbf{\tilde{S}_{\it{j}}}
    %            + \mathbf{D_{\it{ij}}} \cdot \mathbf{\tilde{S}_{\it{i}}} \times \mathbf{\tilde{S}_{\it{j}}}
                 + \mathbf{\tilde{S}_{\it{i}}} \cdot \mathbf{\Gamma}_{ij} \cdot \mathbf{\tilde{S}_{\it{j}}},
\label{eq_spin_H}
\end{equation}
where $\mathbf{{\it J}}_{ij}$ is the isotropic Heisenberg exchange and $\mathbf{\Gamma}_{ij}$ is a symmetric traceless
second-rank tensor that describes the symmetric exchange anisotropy.
Considering the three-fold rotational symmetry around the M-M link, it is convenient to have one of the coordinates along
the line defined by the two M sites.
%We use the local frame $\{xyz\}$ as indicated in Fig.\,\ref{material_model}, with both Ir ions on the $z$ axis.
%In this local frame, the $\bar{ \bar {\mathbf{\Gamma}}}$ tensor is diagonal:
We therefore use the local frame indicated in Fig.\,\ref{material_model}, with both Ir ions on the $z$ axis. In this coordinate system
the $\bar{ \bar {\mathbf{\Gamma}}}$ tensor is diagonal and, for symmetry reasons, can be written as
%
%\begin{equation}
%\bar{ \bar {\mathbf{\Gamma}}} = \begin{pmatrix}
%       {\Gamma}_{xx} &  0             &   0 \\[0.3em]
%         0          &  {\Gamma}_{yy}  &   0\\[0.3em]
%         0          &  0             &    {\Gamma}_{zz} %- {\Gamma}_{xx} - {\Gamma}_{yy}
%     \end{pmatrix},
%\end{equation}
\begin{equation}
\bar{ \bar {\mathbf{\Gamma}}} = \begin{pmatrix}
       \Gamma&  0             &   0 \\[0.3em]
         0          &  \Gamma  &   0\\[0.3em]
         0          &  0             &    -2\Gamma %- {\Gamma}_{xx} - {\Gamma}_{yy}
     \end{pmatrix}.
\end{equation}
%
%with ${\Gamma}_{\!zz}$=$-$(${\Gamma}_{\!xx}$+${\Gamma}_{\!yy}$).
%Then the two-site effective spin Hamiltonian can be written as:
%\begin{equation}
 %{\cal H}_{\langle ij \rangle} =   \mathbf{{\it J}} \mathbf{\tilde{S}_{\it{i}}}\cdot \mathbf{\tilde{S}_{\it{j}}}
 %                +  {\Gamma}_{xx} {\tilde{S}}^{x}_{\it{i}}  {\tilde{S}}^{x}_{\it{j}}
 %               +  {\Gamma}_{yy} {\tilde{S}}^{y}_{\it{i}}  {\tilde{S}}^{y}_{\it{j}}
 %              +  {\Gamma}_{zz} {\tilde{S}}^{z}_{\it{i}} {\tilde{S}}^{z}_{\it{j}}
 %\label{eff_H}
%\end{equation}
%
The eigenstates of such a two-site ${\tilde{S}}$=1/2 system are
                 the singlet $\lvert\Psi_{\!S}\rangle \!=\!(\lvert \uparrow \downarrow \rangle - \lvert \downarrow \uparrow \rangle)\!/\!\sqrt{2}$ and
the three triplet components $\lvert\Psi_{\!1}\rangle \!=\!(\lvert \uparrow \downarrow \rangle + \lvert \downarrow \uparrow \rangle)\!/\!\sqrt{2}$,
                             $\lvert\Psi_{\!2}\rangle \!=\!(\lvert \uparrow \uparrow \rangle + \lvert \downarrow \downarrow \rangle)\!/\!\sqrt{2}$,
                             $\lvert\Psi_{\!3}\rangle \!=\!(\lvert \uparrow \uparrow \rangle - \lvert \downarrow \downarrow \rangle)\!/\!\sqrt{2}$.
The corresponding eigenvalues are
%
%\begin{equation}
%\begin{aligned}
%  &    E_{\!S}= - \frac{3}{4}J,  \hspace{12mm}
%     &E_{1}& = \frac{1}{4}J - \frac{1}{2}{\Gamma}_{yy},         \\
%  &    E_{2} = \frac{1}{4}J - \frac{1}{2}{\Gamma}_{zz},
%      &E_{3}& = \frac{1}{4}J - \frac{1}{2}{\Gamma}_{xx}.
%\end{aligned}
%\label{eq_eigenstates}
%\end{equation}
\begin{equation}
\begin{aligned}
  &    E_{\!S}= - \frac{3}{4}J,  \hspace{12mm}
     &E_{1}& = \frac{1}{4}J - \frac{1}{2}{\Gamma},         \\
  &    E_{2} = \frac{1}{4}J +  {\Gamma},
      &E_{3}& = \frac{1}{4}J - \frac{1}{2}{\Gamma}.
\end{aligned}
\label{eq_eigenstates}
\end{equation}
Expression (\ref{eq_spin_H}) can be then simplified to
\begin{equation}
 {\cal H}_{ij} =   \mathbf{{\it {\bar J}}} \mathbf{\tilde{S}_{\it{i}}}\cdot \mathbf{\tilde{S}_{\it{j}}}
                 + {\bar{\Gamma}} {\tilde{S}}^{z}_{\it{i}} {\tilde{S}}^{z}_{\it{j}},
 \label{eq_re_eff_H}
\end{equation}
where $\mathbf{{\it {\bar J}}} \equiv {\it J} + {\Gamma}$ and ${\bar{\Gamma}} \equiv -3{\Gamma}$.

The first step in the actual QC calculations is defining a relevant set of Slater determinants  in the prior complete-active-space self-consistent-field (CASSCF) treatment \cite{book_QC_00}.
For two IrO$_{6}$ (RhO$_{6}$) octahedra, an optimal choice is having five electrons and three ($t_{2g}$) orbitals
at each of the two magnetically active Ir (Rh) sites. The self-consistent-field optimization was carried out for an average of the lowest nine singlet and
lowest nine triplet states associated with this manifold. Subsequent multireference configuration-interaction (MRCI) computations were performed for each spin multiplicity,
either singlet or triplet, as nine-root calculations.
All these states entered the spin-orbit treatment \cite{SOC_molpro}, in both CASSCF and MRCI.
Within the group of 36 spin-orbit eigenvectors associated with the $t^{5}_{2g}\!-\!t^{5}_{2g}$ manifold,
the lowest-lying four \enquote{magnetic}
states are separated by a significant energy gap from the other 32 states.
The latter correspond to on-site $j\!\approx\!$ 3/2 to $j\!\approx\!$ 1/2 transitions, and are therefore
left aside in the actual mapping procedure. In other words, given the strong SOC and large $j\!\approx\!$ 3/2 to $j\!\approx\!$ 1/2
excitation energies, the initial 36$\times$36 problem can be smoothly mapped onto a 4$\times$4 construction as defined by the
effective Hamiltonian (\ref{eq_spin_H}).
%\cite{IrO_mott_kim_08}.
%

All computations were carried out with the {\sc molpro} quantum-chemistry software \cite{Molpro12}.
In the MRCI treatment, single and double excitations from
the six Ir (Rh) $t_{2g}$ orbitals and from the $2p$ shells of the bridging O ligand sites were taken into account.
The Pipek-Mezey localization module \cite{Pipek89} available in {\sc molpro}
was employed for separating the metal 5$d$ (4$d$) and O 2$p$ valence orbitals into different groups.
To derive the magnitude of direct exchange, we additionally performed calculations
%restricted open-shell Hartree-Fock (rAS) calculations \cite{book_QC_00}.
%A minimal $t_{2g}$-only active space was used in the latter and no intersite excitations were allowed.
in which the active space is again defined by ten electrons and six orbitals but intersite $t_{2g}$$-$$t_{2g}$
excitations are forbidden by restricting to maximum five the number of electrons per TM site. We refer to these
results as rAS (restricted active space, maximum one hole per site).

%
%%%%%%%%%%%%%%
%% FIGURE 2 %%
%%%%%%%%%%%%%%
\begin{figure*}[t]
 \centering
 \hspace{-0.5cm}
 \subfigure{
\raisebox{-0.5cm}{\textbf{ \large (a)}} \hspace{-0.5cm}
 \includegraphics[width=0.275\textwidth,angle=-90]{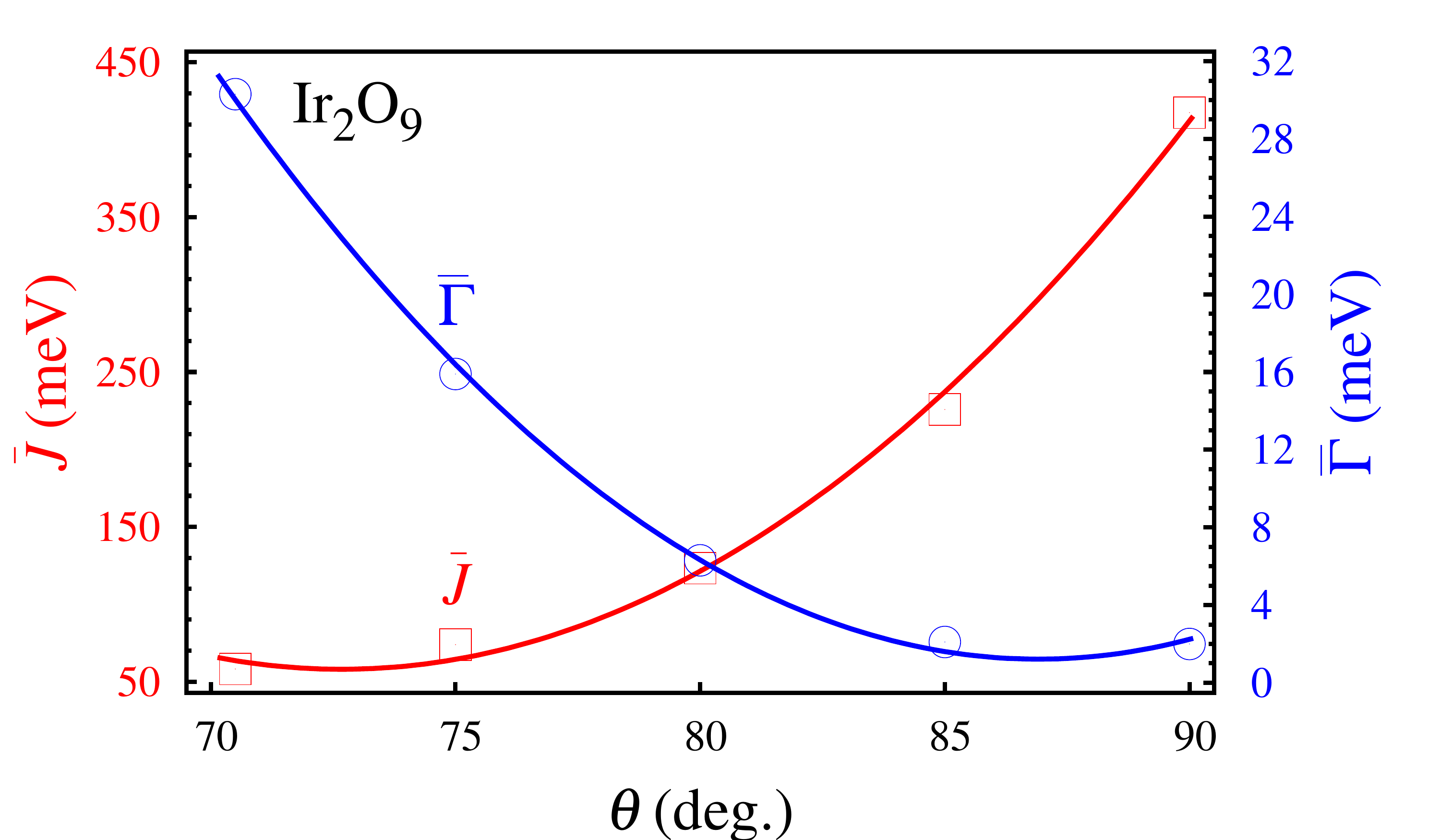}
\label{J_Ir_angle}}
 \quad
 \subfigure{
\raisebox{-0.5cm}{\textbf{ \large (b)}} \hspace{-0.5cm}
 \includegraphics[width=0.275\textwidth,angle=-90]{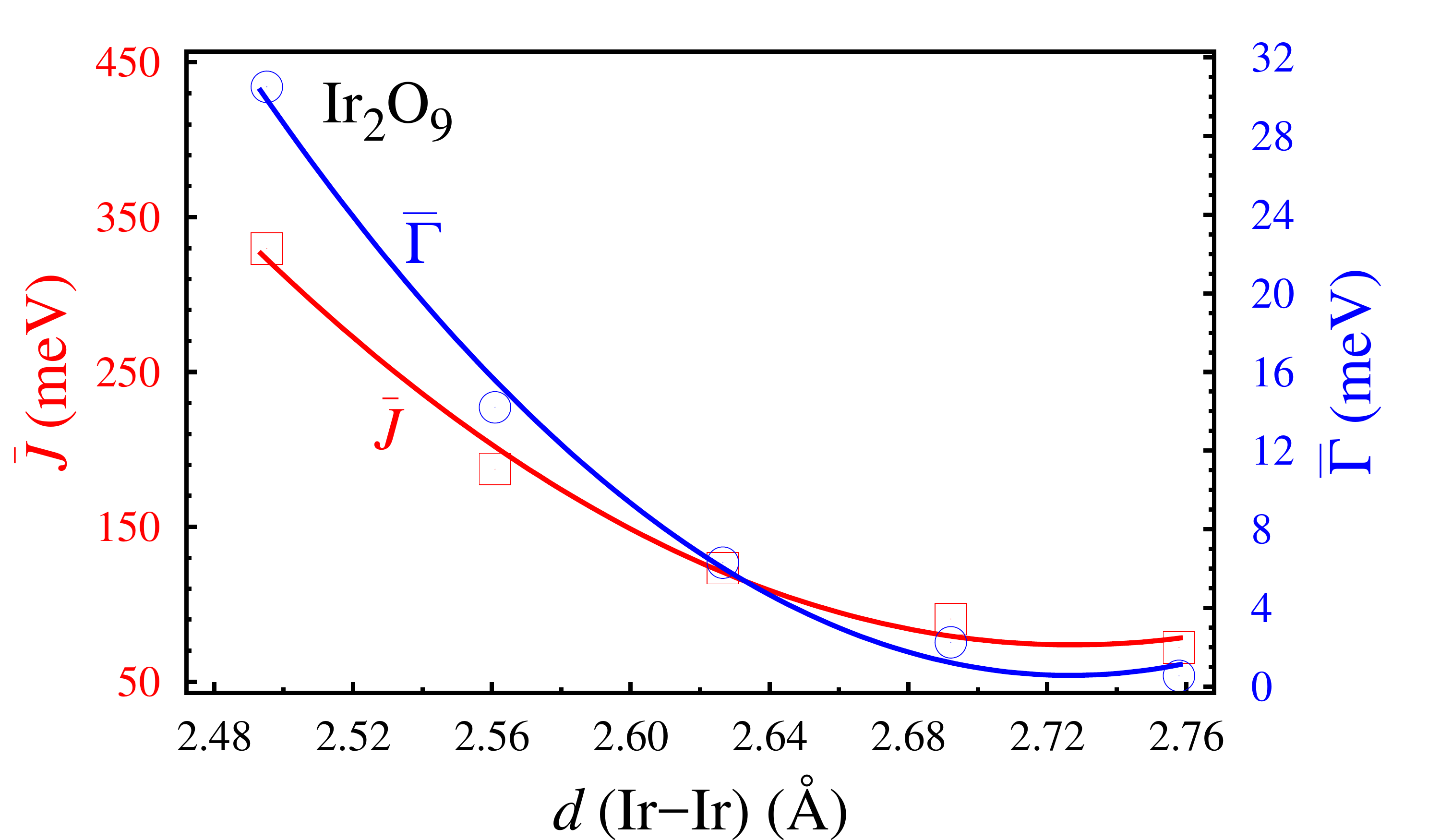}
\label{J_Ir_dist}}
 \vspace{-0.2cm}
 \hspace{-0.8cm}
 \quad
 \subfigure{
 \raisebox{-0.5cm}{\textbf{ \large (c)}} \hspace{-0.5cm}
 \includegraphics[width=0.275\textwidth,angle=-90]{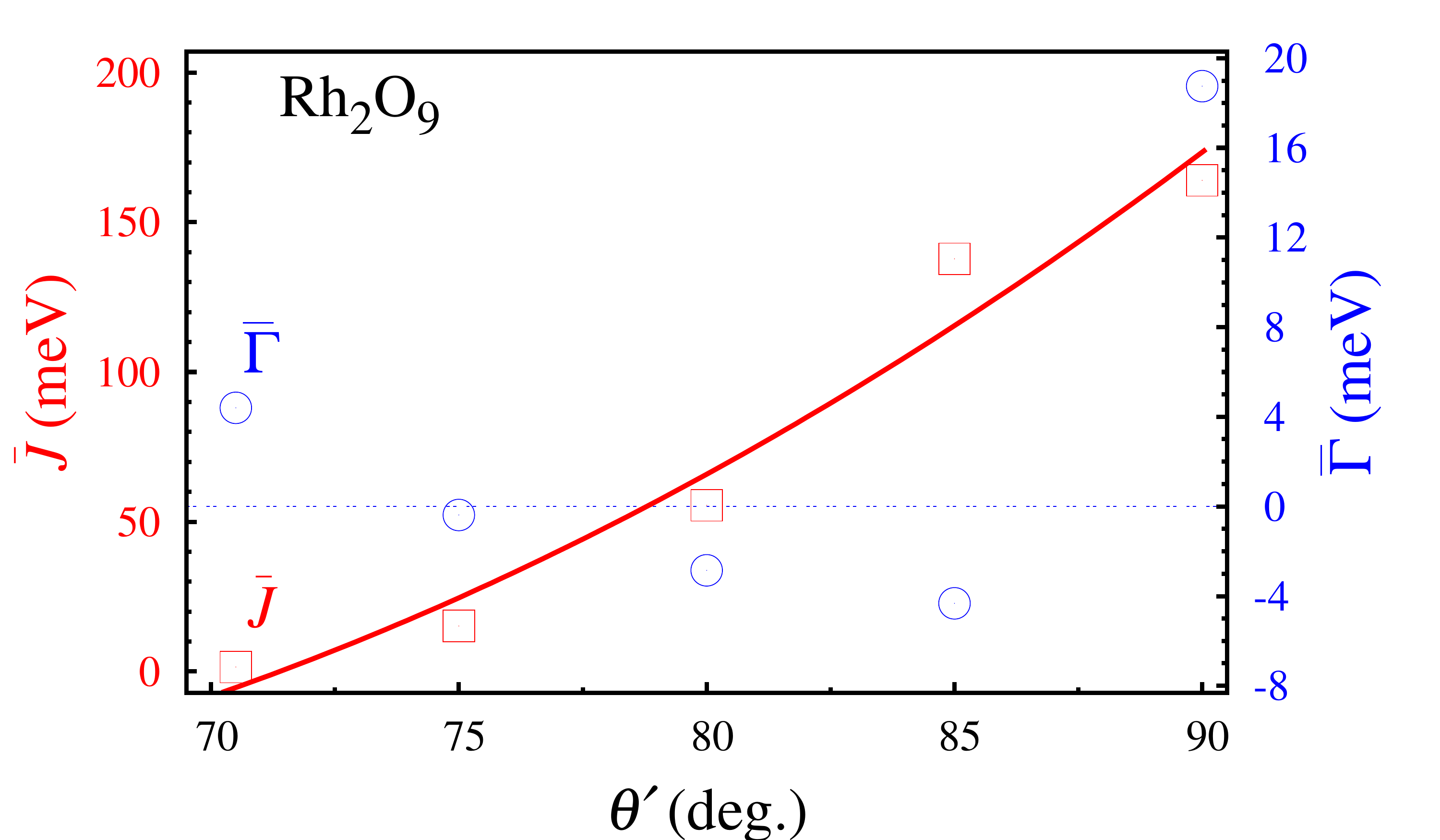}
\label{J_Rh_angle}}
 \quad
 \subfigure{
\raisebox{-0.5cm}{\textbf{ \large (d)}} \hspace{-0.5cm}
 \includegraphics[width=0.28\textwidth,angle=-90]{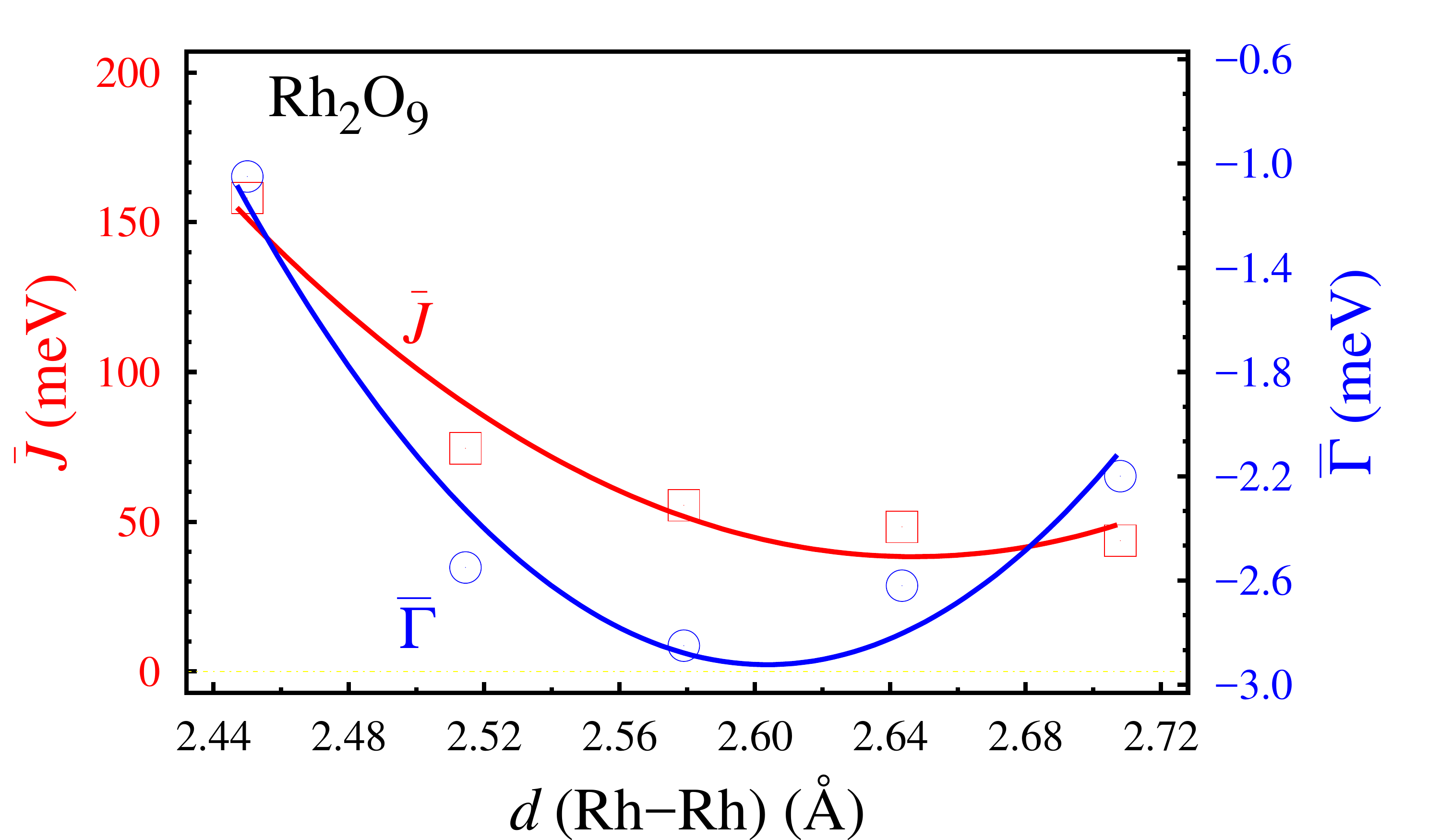}
\label{J_Rh_dist}}
 \caption{
Dependence on structural details of the NN effective magnetic couplings for Ir$_2$O$_9$ and Rh$_2$O$_9$ fragments
of face-sharing octahedra, spin-orbit MRCI results.
(a, c) Variations as function of the Ir-O-Ir bond angle $\theta$ and Rh-O-Rh bond angle $\theta^{\prime}$ when fixing the Ir-Ir distance to $d_0\!=\!2.63$ \AA \ and the Rh-Rh
distance to $d^{\prime}_0\!=\!2.58$ \AA \ . Ligands are radially displaced in planes perpendicular to the $z$ axis. 
Curves were drawn as a guide to the eye; a simple fit is not possible for $\bar \Gamma$ in the case of $4d$ magnetic sites.
(b, d) Variations as function of the TM-TM interatomic distance
when keeping unchanged the positions of the ligands.
%The relative distances from the ligands to the $z$ axis are $r_0\!=\!1.57$ \AA \ for Ir$_2$O$_9$ and $r^{\prime}_0\!=\!1.54$ \AA \ for Rh$_2$O$_9$.
The latter are at distances of $r_0\!=\!1.57$ \AA \ (Ir$_2$O$_9$) and $r^{\prime}_0\!=\!1.54$ \AA \ (Rh$_2$O$_9$) from the $z$ axis.
}\label{coupling}
\end{figure*}

%{\bf{ $[$Ir$_{2}$O$_{9}$$]$ unit.}}
%$\bm{[}${\bf Ir}$\bm{_2}${\bf O}$\bm{_9}$$\bm{]}$ {\bf unit.}

\subsection{ $\bm{[}$Ir$_{2}$O$_{9}$$\bm{]}$ unit}
Effective magnetic couplings for Ir$_2$O$_9$ fragments of two face-sharing IrO$_6$ octahedra
are listed in Table\;\ref{Ir_coupling_80dre}, for an Ir-Ir interatomic distance $d_0\!=\!2.63$ \AA \ and ligand
coordinates that provide Ir-O-Ir angles $\theta_0\!=\!80^{\circ}$.
These structural parameters, obtained by averaging the bond lengths and bond angles in the
experimetally determined lattice configuration of BaIrO$_3$ \cite{Stitzer_BaRhO3_2004}, correspond to slightly elongated octahedra.
For cubic (undistorted) octahedra, $\theta_c\!=\!70.52^{\circ}$.

Results at three different levels of approximation are shown: spin-orbit rAS (rAS+SOC), CASSCF (CAS+SOC), and MRCI (CI+SOC).
%Comparison between these three sets of reported results allows one to estimate roughly relative strengths of three principal contributions  to the exchange mechanism. These are (a)  FM potential exchange owing to a direct overlap of NN metal $d$ orbitals, which is usually small in TM oxides, (b) AFM kinetic exchange due to the direct $d-d$ electron/hole hopping and (c) AFM superexchange due to inderect hopping via  bridging oxygens; in both  (b) and (c) processes intermediate polar states are involved.
%
%%%%%%%%%%%%%
%% TABLE I %%
%%%%%%%%%%%%%
\begin{table}[!t]
\caption{
Energy splittings for the lowest four spin-orbit states of two face-sharing NN IrO$_{6}$ octahedra and the corresponding effective
coupling constants at different levels of approximation, for $d_0\!=\!2.63$ \AA \ and $\theta_{0}\!=\!80^{\circ}$ (all values in meV).
The $J$ values without SOC by rAS, CAS, and CI are $-$1.2, 27.4, and 35.4 meV, respectively.
}
\begin{tabular}{cccc}
\hline
\hline\\[-0.30cm]
 \hspace{-0mm}   \hspace{0mm}  &\hspace{0mm} rAS+SOC \hspace{0mm} &\hspace{0mm} CAS+SOC  \hspace{-0mm} &CI+SOC\\%%($\times\!2$)\\
\hline
\\[-0.20cm]
$\Psi_{\!S}\!=\!(\lvert \uparrow \downarrow \rangle - \lvert \downarrow \uparrow \rangle)\!/\!\sqrt{2}$         &15.2         & 0.0        &   0.0  \\[0.1cm]
$\Psi_{\!2}\!=\!(\lvert \uparrow \uparrow \rangle + \lvert \downarrow \downarrow \rangle)\!/\!\sqrt{2}$         & 0.4         &72.0        & 123.3  \\[0.1cm]
$\Psi_{\!3}\!=\!(\lvert \uparrow \uparrow \rangle - \lvert \downarrow \downarrow \rangle)\!/\!\sqrt{2}$         & 0.0         &74.0        & 126.5  \\[0.1cm]
$\Psi_{\!1}\!=\!(\lvert \uparrow \downarrow \rangle + \lvert \downarrow \uparrow \rangle)\!/\!\sqrt{2}$         & 0.0         &74.0        & 126.5  \\[0.1cm]
${\bar J},{\bar \Gamma}$                                                                                             &$-$14.9, $-$0.7    &72.0, 4.1      &123.3, 6.3  \\[0.1cm]
\hline
\hline
\end{tabular}
\label{Ir_coupling_80dre}
\end{table}
%
%%%%%%%%%%%%%
%% TABLE II%%
%%%%%%%%%%%%%
%%\begin{table}[!b]
%%\caption{
%%Energy splittings for the four lowest spin-orbit states of two face-sharing NN IrO$_{6}$ octahedra and the corresponding effective
%%coupling constants at different levels of approximation, for $d_0\!=\!2.63$ \AA \ and $\theta_{0}\!=\!90^{\circ}$ (all in meV).
%%The $J$ values without SOC by rAS, CAS, and CI are 0.04, 64.7, and 116.4 meV, respectively.
%%}
%%\begin{tabular}{cccc}
%%\hline
%%\hline\\[-0.30cm]
 %%\hspace{-0mm} States \hspace{0mm}  &\hspace{0mm} rAS+SOC \hspace{0mm} &\hspace{0mm} CAS+SOC  \hspace{-0mm} &CI+SOC\\%%($\times\!2$)\\
%%\hline
%%\\[-0.20cm]
%%$\Psi_{\!S}\!=\!(\lvert \uparrow \downarrow \rangle - \lvert \downarrow \uparrow \rangle)\!/\!\sqrt{2}$         &17.0         & 0.0        &   0.0  \\[0.1cm]
%%$\Psi_{\!2}\!=\!(\lvert \uparrow \uparrow \rangle + \lvert \downarrow \downarrow \rangle)\!/\!\sqrt{2}$        & 0.0          & 293.6        & 417.5  \\[0.1cm]
%%$\Psi_{\!3}\!=\!(\lvert \uparrow \uparrow \rangle - \lvert \downarrow \downarrow \rangle)\!/\!\sqrt{2}$         & 0.3          & 294.2        & 418.5  \\[0.1cm]
%%$\Psi_{\!1}\!=\!(\lvert \uparrow \downarrow \rangle + \lvert \downarrow \uparrow \rangle)\!/\!\sqrt{2}$        & 0.3          & 294.2       & 418.5 \\[0.1cm]
%%${\bar J},{\bar \Gamma}$                                                                                             &$-$17.0, $$0.7    &293.6, 1.2      &417.5, 2.0  \\[0.1cm]
%%\hline
%%\hline
%%\end{tabular}
%%\label{Ir_coupling_90dre}
%%\end{table}
%
The rAS data account for only direct $d$$-$$d$ exchange. For $d_0\!=\!2.63$ \AA \ and $\theta_{0}\!=\!80^{\circ}$,
 the rAS ${\bar J}$ is $-$14.9 meV while the anisotropic ${\bar \Gamma}$ is $-$0.7 meV when including SOC.
The magnitude of the ferromagnetic (FM) rAS ${\bar J}$ is similar to that computed in square-lattice 3$d^{9}$ Cu oxides \cite{Guo_1988,Martin_1988,Oosten_1996,Carmen_2002b}
and in the corner-sharing iridate Ba$_{2}$IrO$_{4}$
 \cite{Vamshi_PRX}. The anisotropic ${\bar \Gamma}$ is also FM at the rAS level and its magnitude is slightly larger as compared with the AFM rAS ${\bar \Gamma}$ of
 the corner-sharing iridate Ba$_{2}$IrO$_{4}$ \cite{Vamshi_PRX}.
By CASSCF and MRCI, the singlet $\Psi_{\!S}$ becomes the ground state,
well below the \enquote{triplet} components $\Psi_{\!1}$, $\Psi_{\!2}$, and $\Psi_{\!3}$. This indicates that the
isotropic Heisenberg exchange $\bar J$ ($\bar J\!>\!0$) defines now the largest energy scale. In the CASSCF approximation, only
$t_{2g}$$-$$t_{2g}$ intersite excitations are accounted for, i.e., $t^{6}_{\!2g}$$-$$t^{4}_{\!2g}$ configurations. The $\bar J$ value extracted by CAS+SOC,
72 meV, is twice as large as compared, e.g., to the CASSCF $J$'s in layered 3$d^{9}$ cuprates \cite{Oosten_1996,Guo_1988,Martin_1988,Carmen_2002b}
and in the corner-sharing iridate Ba$_{2}$IrO$_{4}$ \cite{Vamshi_PRX}. In the configuration-interaction treatment,
%which includes $t^{5}_{2g}e^{1}_{g}$-$t^{4}_{2g}$ and O $2p$ to Ir 5$d$ charge-transfer virtual states as well, $\bar J$ is 123.3 meV, about 70\% larger as compared to the CAS+SOC value.
which includes TM $t_{2g}$ to $e_{g}$ and charge-transfer O $2p$ to Ir 5$d$ excitations as well,
%with an extended basis set including the crystal-field $t_{2g}$ to $e_{g}$ on TM as well charge-transfer O $2p$ to Ir $5d$ excitations, the calculated
$\bar J$ is 123.3 meV, about 70\% larger as compared to the CAS+SOC result.
By accounting for correlation effects, the symmetric anisotropic coupling  ${\bar \Gamma}$ is also significantly enlarged, from $-$0.7 meV by rAS+SOC to 6.3 meV by spin-orbit MRCI.

\begin{table}[!b]
\caption{
Energy splittings for the lowest four spin-orbit states and the corresponding effective
coupling constants for variable Ir-O-Ir angle, MRCI+SOC results (meV). The NN Ir-Ir distance is $d_{0}$=2.63 \AA{}.
The Ir-O-Ir angle $\theta$ is listed for each geometry; distances between Ir
and the bridging O's, $d$(Ir-O), are provided within brackets.
}
\hspace{-0.8cm}
\begin{tabular}{c c c c c c}
\hline
\hline\\[-0.25cm]
 \hspace{0mm}  \hspace{0mm}& \hspace{0mm} $70.5^{\circ}$(2.27\AA)  \hspace{0mm}&\hspace{0mm}$75^{\circ}$(2.16\AA) \hspace{0mm}    & \hspace{0mm} $80^{\circ}$(2.04\AA)  \hspace{0mm}
                                                          &\hspace{0mm}$85^{\circ}$(1.94\AA)  \hspace{0mm}&\hspace{0mm} $90^{\circ}$(1.86\AA) \\
\hline
\\[-0.20cm]
    $\Psi_{\!S}$      &   0.0             &   0.0    & 0.0      &   0.0          &   0.0 \\
    $\Psi_{\!2}$      &  58.2             &  73.9   & 123.3       & 225.8          & 417.5 \\
    $\Psi_{\!1}$      &  73.4             &  81.8   & 126.5       & 226.9          & 418.5 \\
    $\Psi_{\!3}$      &  73.4             &  81.8    & 126.5      & 226.9          & 418.5  \\
${\bar J},{\bar \Gamma}$  &  58.2, 30.3      &  73.9, 15.9   & 123.3, 6.3  & 225.8, 2.1   & 417.5, 2.0  \\[0.1cm]
\hline
\hline
\end{tabular}
\label{Ir_coupling_angl}
\end{table}

In the case of face-sharing ligand octahedra, the TM ions often form dimers, trimers, or chains \cite{Stitzer_BaRhO3_2004}. This type of low-dimensional packing usually results in sizable distortions of the ligand cages.
It is known that the  effective spin interactions are strongly dependent on structural details such as bond angles \cite{Vamshi_2016nc,Vamshi_Li2RhO3,Ravi_RuCl3,Ravi_2017arxiv} and bond lengths \cite{Ravi_2018arxiv}.
For better insight into the dependence of the NN magnetic couplings on such structural parameters, we performed additional calculations for distorted geometries with all ligands pushed
closer to (or farther from) the Ir-Ir axis, which therefore yields larger (or smaller) Ir-O-Ir bond angles while keeping the overall $D_{3h}$ point-group symmetry.
The resulting MRCI+SOC data are provided in
Table\;\ref{Ir_coupling_angl}. The overall trends for the magnetic couplings $\bar J$ and ${\bar \Gamma}$ are illustrated graphically in Fig.\;\ref{J_Ir_angle}. It is
seen that the angle dependence for both $\bar J$ and ${\bar \Gamma}$ can be rather well reproduced with parabolic curves. The Heisenberg $\bar J$ displays a steep increase with
larger angle, i.e., from 58 meV at 70.5$^{\circ}$ to 417 meV at 90$^{\circ}$. On the other hand, the anisotropic coupling ${\bar \Gamma}$
shows a rapid decrease, from a remarkably large value of 30 meV at 70.5$^{\circ}$  to 2 meV at 90$^{\circ}$.

We further analyzed the dependence on the Ir-Ir interatomic distance $d$(Ir-Ir) of the magnetic interactions.
%In this additional set of calculations, the distances between the ligands and the midpoint of the Ir-Ir bond were fixed to 1.57 \AA{},
In this set of calculations, the distance between the O ligands and the $z$ axis (along the Ir-Ir bond) was fixed to 1.57 \AA{},
while $d$(Ir-Ir) was either increased or reduced by
up to 5$\%$ with respect to the reference Ir-Ir separation $d$=$d_{0}$=2.63 \AA.
As shown in Fig.\;\ref{J_Ir_dist} (see also Appendix B, Table\;\ref{App_Ir_dist}), both $\bar J$
and ${\bar \Gamma}$ have again pronounced parabolic dependence on $d$(Ir-Ir). In contrast to the
variations as function of angle displayed in Fig.\;\ref{J_Ir_angle}, here $\bar J$
and ${\bar \Gamma}$ follow the same trend. More specifically, both $\bar J$ and ${\bar \Gamma}$ rapidly increase with decreasing $d$(Ir-Ir).

We also performed calculations in which the six O ligands not shared by the Ir ions were displaced as well along the $z$ axis,
such that each Ir site remains in the center of the respective octahedron. We found that the differences between the ${\bar J}$ values obtained from these computations
and the corresponding ${\bar J}$'s in Fig.\;\ref{J_Ir_dist} are rather small, not more than 15\%.

\begin{figure}[b]
 \centering
 \vspace{-0.0cm}
 \subfigure{
\raisebox{4.2cm}{\textbf{ \large (a)}} %\hspace{0.4cm}
\includegraphics[width=0.182\textwidth]{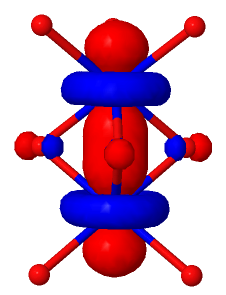}
\label{a1g_cd}
}
 \quad
\subfigure{
\raisebox{4.2cm}{\textbf{ \large (b)}}
 \includegraphics[width=0.182\textwidth]{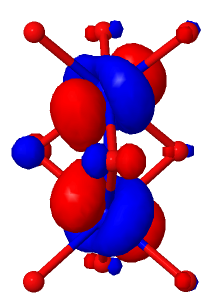}
\label{e1g_cd}}
 \caption{
% Ir 5$d$ a$_{1g}$ orbitals charge distribution as obtained by CASSCF for Ir$_{2}$O$_{9}$ fragment of face-sharing octahedra. The
%a$_{1g}$ orbitals are almost $d_{z^{2}}$ character with the lobes directing through the common face.
Natural orbitals of $a_{1g}$ (a) and $e^{\pi}_{\!g}$ (b) type for a Ir$_{2}$O$_{9}$ fragment of face-sharing octahedra, as obtained by CASSCF calculations.
The former have strong $\sigma$-type overlap.
%(a) The $a_{1g}$ orbitals are almost $d_{z^{2}}$ character with the lobes directing through the common face.
%(b) The $\sigma$-like symmetry adapted $e^{\pi}_{\!g}$ combinations of bridging O $2p$ orbitals have large overlap with the Ir $5d$ functions.
}
\label{a1g_e1g_cd}
\end{figure}

%\textcolor{blue} {%The field generated by octahedral ligand cages splits the $d$ states into $t_{2g}$ and $e_g$ levels.
The face-sharing linkage and additional distortions applied to the two-octahedra clusters
%lower the symmetry of the 2-octahedra units to $D_{3h}$, leading to further
split the $t_{2g}$ levels into $a_{1g}$ and $e^{\pi}_{g}$ components.
For all Ir$_{2}$O$_{9}$ units considered here, we find that the $a_{1g}$ sublevels lie at lower energy and
that the $t_{2g}$ hole has $e_{g}^{\pi}$ character without accounting for SOC.
%For all Ir$_{2}$O$_{9}$ units considered here, we find that the $a_{1g}$ component lies at lower energy, and
%hence the $t_{2g}$ hole in the ground-state $t^{5}_{2g}$ configuration has $e_{g}^{\pi}$ character.
%
The $a_{1g}$ orbitals belonging to NN sites have substantial direct overlap [see Fig\,\ref{a1g_cd}],
much larger than in the case of $e_{g}^{\pi}$ orbitals [see Fig\,\ref{e1g_cd}].
The rather small AFM Heisenberg $J$ derived from the calculations without SOC (see caption of Table \ref{Ir_coupling_80dre}) is therefore the result of
(relatively) weak direct exchange involving the higher-lying $e_{g}^{\pi}$ states.
By accounting for spin-orbit interactions, however, the Heisenberg $J$ is enhanced to impressive values that are up to three
times larger than the results obtained without SOC (72 vs 27 meV at the CASSCF level, 123 vs 35 meV by MRCI, see Table \ref{Ir_coupling_80dre}).
This strong increase of the Heisenberg $J$ is the consequence of mixing $a_{1g}$ character to the spin-orbit ground-state wave-function.
%}
%This strong increase of the Heisenberg $J$ is the consequence of admixing $a_{1g}$ orbital in the entangled spin-orbit ground-state wave-function.}
%
% For the $5d$ orbitals of Ir$^{4+}$ ions in $D_{3h}$ environment, the $t_{2g}$ levels are split into $a_{1g}$
% and $e^{\pi}_{g}$ components.
% %The local distortions of the face-sharing octahedra lead to the splitting of $t_{2g}$ orbitals into the $a_{1g}$ singlet and the $e_{g}^{\pi}$ doublet.
%  For all Ir$_{2}$O$_{9}$ units considered here, we find that the $a_{1g}$ component lies at lower energy and that
%  the $t_{2g}$ hole has $e_{g}^{\pi}$ character.
%  %
%  When SOC is not considered, the AFM Heisenberg $J$ comes from the $d$-$d$ overlap between $e_{g}^{\pi}$ orbitals.
%  %
%  If SOC is further taken into account within the $t_{2g}$ manifold, since the $t_{2g}$ splitting is comparable to the SOC,
%  the $j_{\!e\!f\!f}$=1/2 states mix with the $j_{\!e\!f\!f}$=3/2 states.
%  In this case,  in addition to the contribution of the AF $J_{e_{g}^{\pi}}$ from the overlap between $e_{g}^{\pi}$ orbitals,
%  the contribution of AF $J_{a_{1\!g}}$ from the
%  large direct $d$-$d$ overlap between $a_{1g}$ orbitals (see Appendix Fig.\;\ref{a1g_cd})
%  becomes significant, thus gives larger AF SOC $\bar J$'s.
%
%
%%%%%%%%%%%%%
%%%    TABLE III   %%%
%%%%%%%%%%%%%
\begin{table}[!b]
\caption{
Energy splittings for the lowest four spin-orbit states of two face-sharing NN RhO$_{6}$ octahedra and the corresponding effective
coupling constants at different levels of approximation, for $d^{\prime}_{0}\!=\!2.58$ \AA \  and $\theta_{0}^{\prime}\!=\!80^{\circ}$ (all in meV).
The $J$ values without SOC by rAS, CAS, and CI are $-$0.9, 19.0, and 29.4 meV, respectively.
}
\begin{tabular}{cccc}
\hline
\hline\\[-0.30cm]
 \hspace{-0mm}  \hspace{0mm}  &\hspace{0mm} rAS+SOC \hspace{0mm} &\hspace{0mm} CAS+SOC  \hspace{-0mm} &CI+SOC\\%%($\times\!2$)\\
\hline
\\[-0.20cm]
$\Psi_{\!S}\!=\!(\lvert \uparrow \downarrow \rangle - \lvert \downarrow \uparrow \rangle)\!/\!\sqrt{2}$         &16.0         & 0.0        &   0.0  \\[0.1cm]
$\Psi_{\!3}\!=\!(\lvert \uparrow \uparrow \rangle - \lvert \downarrow \downarrow \rangle)\!/\!\sqrt{2}$         & 0.0          &25.4        & 54.0  \\[0.1cm]
$\Psi_{\!1}\!=\!(\lvert \uparrow \downarrow \rangle + \lvert \downarrow \uparrow \rangle)\!/\!\sqrt{2}$         & 0.0         &25.4        & 54.0  \\[0.1cm]
$\Psi_{\!2}\!=\!(\lvert \uparrow \uparrow \rangle + \lvert \downarrow \downarrow \rangle)\!/\!\sqrt{2}$         & 1.1         &26.5        & 55.5  \\[0.1cm]
${\bar J},{\bar \Gamma}$                                                                                                   &$-$14.9,$-$2.3    &26.5,$-$2.2      &55.5,$-$2.8  \\[0.1cm]
\hline
\hline
\end{tabular}
\label{Rh_coupling_80dre}
\end{table}
%
%%%%%%%%%%%%%
%% TABLE IIII%%
%%%%%%%%%%%%%
%%\begin{table}[!b]
%%\caption{
%%Energy splittings for the four lowest spin-orbit states of two face-sharing NN RhO$_{6}$ octahedra and the corresponding effective
%%coupling constants, MRCI+SOC results (meV). The NN Rh-Rh distance is 2.58 \AA{}. The Rh-O-Rh angle $\theta_{0}^{\prime}$ and the
%%corresponding distance between Rh and the bridging O's, $d$(Rh-O), are listed for each geometry.
%%}
%%\hspace{-2mm}
%%\begin{tabular}{ccccc}
%%\hline
%%\hline\\[-0.25cm]
%% \hspace{4mm}States \hspace{1mm}& \hspace{1.5mm} $70.5^{\circ}$,2.23\AA  \hspace{1.5mm}&\hspace{1.5mm}$75^{\circ}$,2.12\AA \hspace{1mm}&\hspace{1.5mm}$85^{\circ}$,1.91\AA  \hspace{1.5mm}&\hspace{1.5mm} $90^{\circ}$,1.82\AA \\
%%\hline
%%\\[-0.20cm]
%%    $\Psi_{\!S}$      &   0.0             &   0.0          &   0.0           &   0.0 \\
%%    $\Psi_{\!3}$      &   3.7             &  15.0         & 135.7         & 173.4 \\
%%    $\Psi_{\!1}$      &   3.7             &  15.0         & 135.7         & 173.4 \\
%%    $\Psi_{\!2}$      &   1.5             &  15.2         & 137.8         & 164.0  \\
%%${\bar J},{\bar \Gamma}$  &  1.5, 4.4       &  15.2, $-$0.4    & 137.8, $-$4.3   & 164.0, 18.8  \\[0.1cm]
%%\hline
%%\hline
%%\end{tabular}
%%\label{Rh_coupling_angl}
%%\end{table}

%
%
%%%%%%%%%%%%%
%% TABLE IIII%%
%%%%%%%%%%%%%
\begin{table}[!t]
\caption{
Energy splittings and the corresponding effective coupling constants for variable Rh-O-Rh angle,
MRCI+SOC results (meV). The NN Rh-Rh distance is 2.58 \AA{}.
%The Rh-O-Rh angle $\theta_{0}^{\prime}$ and the
%corresponding distance between Rh and the bridging O's, $d$(Rh-O), are listed for each geometry.
The Rh-O-Rh angle $\theta^{\prime}$ is listed for each geometry; distances between Rh
and the bridging O's, $d$(Rh-O), are provided within brackets.
}
%\centering
\hspace{-0.0cm}
\begin{tabular}{c c c c c c}
\hline
\hline\\[-0.25cm]
 \hspace{0mm}  \hspace{0mm}& \hspace{0mm} $70.5^{\circ}$(2.23\AA)  \hspace{0mm}&\hspace{0mm}$75^{\circ}$(2.12\AA) \hspace{0mm} &\hspace{0mm}$80^{\circ}$(2.04\AA) \hspace{0mm}
  							&\hspace{0mm}$85^{\circ}$(1.91\AA)  \hspace{0mm}&\hspace{0mm} $90^{\circ}$(1.82\AA) \\
\hline
\\[-0.20cm]
    $\Psi_{\!S}$      &   0.0             &   0.0     &  0.0     &   0.0           &   0.0 \\
    $\Psi_{\!3}$      &   3.7             &  15.0    & 54.0     & 135.7         & 173.4 \\
    $\Psi_{\!1}$      &   3.7             &  15.0     & 54.0    & 135.7         & 173.4 \\
    $\Psi_{\!2}$      &   1.5             &  15.2      & 55.5   & 137.8         & 164.0  \\
${\bar J},{\bar \Gamma}$  &  1.5, 4.4     &  15.2,$-$0.4    & 55.5,$-$2.8  & 137.8,$-$4.3 & 164.0, 18.8  \\[0.1cm]
\hline
\hline
\end{tabular}
\label{Rh_coupling_angl}
\end{table}

%$\bm{[}${\bf Rh}$\bm{_2}${\bf O}$\bm{_9}$$\bm{]}$ {\bf unit.}
\subsection{ $\bm{[}$Rh$_{2}$O$_{9}$$\bm{]}$ unit}

In order to make a informative comparison between 5$d$ and 4$d$ oxides, we also performed calculations for the effective magnetic couplings on [Rh$_{2}$O$_{9}$] fragments consisting
of two face-sharing RhO$_{6}$ octahedra, with a Rh-Rh interatomic distance $d^{\prime}_{0}\!=\!2.58$ \AA \ and Rh-O-Rh bond angles $\theta_{0}^{\prime}\!=\!80^{\circ}$.
As for the material model of face-sharing 5$d^{5}$ octahedra, the structural parameters of the Rh$_{2}$O$_{9}$ cluster were chosen according to the average bond lengths and bond angles
of the BaRhO$_{3}$ compound. We used in this regard the crystallographic data reported in Ref. \cite{Stitzer_BaRhO3_2004}.
QC results are presented in Table\;\ref{Rh_coupling_80dre}. Interestingly, the $\bar J$ value obtained by spin-orbit rAS is
the same as for the Ir$_{2}$O$_{9}$ cluster (Table\;\ref{Ir_coupling_80dre}).
However, the $\bar J$'s obtained by CAS+SOC and CI+SOC are significantly smaller as compared with those in Table\;\ref{Ir_coupling_80dre}.
Still, $\bar J$ remains much larger than the magnetic couplings in the edge-sharing 4$d^{5}$ compounds Li$_{2}$RhO$_{3}$ and $\alpha$-RuCl$_{3}$ \cite{Vamshi_Li2RhO3,Ravi_RuCl3}.
%It is  also worth pointing out that ${\bar \Gamma}$ is always FM in these calculations.

Energy splittings within the group of the four low-lying $d^{5}\!-d^{5}$ states and the resulting effective
coupling constants for different Rh-O-Rh angles are listed in Table\;\ref{Rh_coupling_angl}. Furthermore, the dependence of $\bar J$ and ${\bar \Gamma}$ on the Rh-O-Rh
bond angles and on the Rh-Rh interatomic distances are illustrated in Fig.\;\ref{J_Rh_angle} and Fig.\;\ref{J_Rh_dist}, respectively (for more details see Appendix B, Table\;\ref{App_Rh_angl}
and Table\;\ref{App_Rh_dist}). As indicated in Fig.\;\ref{J_Rh_angle}, $\bar J$ displays nearly linear behavior with variable angle, increasing from 1.5 meV at 70.5$^{\circ}$ to 164 meV
at 90$^{\circ}$. ${\bar \Gamma}$ changes sign from AFM to FM coupling close to 75$^{\circ}$, with a minimum of $-$6 meV at $\!85^{\circ}$, and then
changes back to AFM values for larger angles. On the other hand, with variable $d$(Rh-Rh) [Fig.\;\ref{J_Rh_dist}], ${\bar \Gamma}$ is always FM, with a minimum
of $-$2.8 meV at 2.56 \AA{}, and $\bar J$ features a similar trend as for Ir sites in Fig.\;\ref{J_Ir_dist}.

%
%%%%%%%%%%%%%%
%% FIGURE 1 %%
%%%%%%%%%%%%%%
\section{Discussion}

We analyze in more detail in this section the relative values of the different contributions to intersite exchange, i.e., direct $t_{2g}$$-$$t_{2g}$ exchange,
$t_{2g}$$-$$t_{2g}$ electron/hole hopping, and indirect hopping via the bridging oxygens.
In first place, it is clear that a systematically small portion of FM potential exchange to the overall $\bar J$ is here of secondary importance.
The contribution coming from direct  hopping can be straightforwardly
estimated from the CASSCF $J$ since only
intersite M($t_{2g}$)--M($t_{2g}$) excitation processes ($t^6_{2g}$--$t^4_{2g}$ polar configurations) are taken into account at the CASSCF level.
In the CI treatment, superexchange paths including the bridging-ligand $2p$ and TM $e_{g}$ orbitals are also added on top of direct hopping, providing
a more comprising description of intersite exchange mechanisms.
In the case of BaIrO$_3$, for instance, when the Ir-O-Ir bond angle is $80^{\circ}$, the exchange calculated at the CASSCF level (without SOC),
$J$ = 27.4 meV,  is already  77$\%$ of the CI result, 35.4 meV (see Table\;\ref{Ir_coupling_80dre}). While this fraction is significantly reduced
if the Ir-O-Ir bond angle is modified towards $90^{\circ}$ (see Table\;\ref{App_Ir_angl}), indicating that the $d-p-d$ superexchange contribution starts to rise
as a result of shorter Ir-O bonds, 
%It also increases when switching on SOC (Table\;\ref{Ir_coupling_80dre}).
the data computed for 80$^{\circ}$ bond angles show that, given the large direct-hopping integrals, the direct AFM $d$$-$$d$ superexchange
may surpass the $d$$-$$p$$-$$d$ superexchange.
The two mechanisms should be considered in any case on equal footing for high-quality estimates.
In the context of recent discussions on the role of the various types of intersite exchange \cite{Kugel_2015,Khomskii_2016},
our QC data provide a more quantitative picture on the different contributions.

\begin{table}[!t]
\caption{
Angle dependence of the trigonal CF splitting $\Delta_t$ (eV) obtained from MRCI calculations
for Ir$_{2}$O$_{9}$ and Rh$_{2}$O$_{9}$ fragments of face-sharing octahedra. The
$a_{1g}$ level is for all angles the lowest in energy.
}
%\hspace{20mm}
\begin{tabular}{ccc}
\hline
\hline\\[-0.30cm]
 \hspace{10mm} $\theta/\theta'$\hspace{10mm}  &\hspace{10mm} $\Delta^{\text {Ir}}_t$ \hspace{10mm}  &\hspace{10mm} $\Delta^{\text {Rh}}_t$ \hspace{10mm} 
  \\
 \hline
\\[-0.10cm]
70.5  & \hspace{0mm} $-$0.86    &   $-$0.69        \\[0.15cm]
75     & \hspace{0mm} $-$0.89    &   $-$0.67         \\[0.15cm]
80     & \hspace{0mm} $-$0.71    &   $-$0.53         \\[0.15cm]
85     & \hspace{0mm} $-$0.43    &   $-$0.29         \\[0.15cm]
90     & \hspace{0mm} $-$0.20    &   $-$0.10          \\[0.15cm]
\hline
\hline
\end{tabular}
\label{t_eff_mod}
\end{table}

Given the facet-sharing geometry, the direct $d$$-$$d$ electron/hole hopping between $a_{1g}$ orbitals is considerable.
This hopping interferes with the indirect hopping via the bridging-oxygen O$_3$ group, providing a total transfer integral $t$.
Since the exchange coupling $J$ is mainly controlled by the square of $t$ ($J \sim t^2/U$, where $U$ is the on-site Coulomb repulsion), a large
$\bar J$ value of up to $\approx$400 meV (see Table \ref{App_Ir_angl} and \ref{App_Ir_dist}) is not surprising.
Both the direct ($\sim$$t^{dd}$) and indirect ($\sim$$t^{dpd}$) transfer  processes can occur trough the $e^{\pi}_{g}$ and $a_{1g}$ channels independently.
As discussed in Appendix A, the total transfer integral $t$ can be then decomposed as $t=w_e t_{e}+ w_a t_a$, where $t_{e,a}=t_{e,a}^{dd} + t_{e,a}^{dpd}$. The corresponding channel weights, $w_e$ and $w_a$,
are controlled by the ratio $\Delta_t/\lambda$, with $\lambda$ being the spin-orbit coupling, 0.47 eV for Ir and 0.15 eV for Rh \cite{RhIr_vmk_IC_14};
the dependence of the trigonal splitting $\Delta_{t}$ on bond angles is illustrated in Table \ref{t_eff_mod}.
The different terms entering the total transfer integral $t$ are expected to behave differently 
when varying the geometry of the M-O$_3$-M structure.
The large direct overlap  between two NN $a_{1g}$ orbitals suggests that the direct hopping $t_{a}^{dd}$ contributes significantly to $t_{a}$, as evidenced in Fig.\,\ref {a1g_e1g_cd}.
In contrast, the $e^{\pi}_{g}$ orbitals are tilted with respect to the $z$ axis,
thus giving rise to weaker direct overlap
and more significant $d$$-$$p$$-$$d$ couplings (see Fig. \ref{e1g_cd}), i.e., a more important role of $t^{dpd}_{e}$ in $t_{e}$.
It is the interplay between these processes, $d-d$ and $d-p-d$ superexchange, that is mainly responsible for the strong
variations as function of bond angles and bond lengths.
From a wider perspective, it is clear that the equilibrium geometrical configuration and
the associated $J$ value depend on interactions and degrees of freedom that also involve the extended
crystalline surroundings. An interesting aspect to be considered is inter-site couplings within the entire M$_{3}$O$_{12}$
block of three face-sharing octahedra along the $c$ axis (see Fig.\ref{crystal_BaIrO3}). One question concerns the possibility
of cooperative M$-$M dimerization as driving force for the charge density wave observed in BaIrO$_{3}$ \cite{Cao_BaIrO3_2000}.
Two-site bond formation on three-center units with a spin 1/2 at each magnetic site and long-range ordering of these {\lq{dimers}\rq}
has been earlier proposed in the quasi-1D system NaV$_{2}$O$_{5}$ \cite{PRL_2002,Hozoi_prb_2003}.

To summarize, we employ quantum chemistry methods
to provide valuable insights on the effective magnetic interactions
in $5d$ and $4d$ oxides with face-sharing oxygen octahedra, BaIrO$_{3}$ and BaRhO$_{3}$.
The same methodology has previously been used to derive
magnetic coupling constants in good agreement with experimental estimates
in the perovskite iridate CaIrO$_{3}$ \cite{Nikolay_CaIrO3_2012,Moretti_2014}, in square-lattice Ba$_{2}$IrO$_{4}$ \cite{Vamshi_PRX}
and Sr$_{2}$IrO$_{4}$ \cite{Ir214_bogdanov_15}, and in pyrochlore iridates \cite{Ravi_2017arxiv}.
The large AFM Heisenberg interactions computed here for face-sharing octahedra are remarkable since they
exceed the values computed so far for corner-sharing \cite{Motoyama_1996,Oosten_1996,Nikolay_CaIrO3_2012,Vamshi_PRX}
and edge-sharing systems \cite{Vamshi_2016nc}.
One peculiar exception with regard to edge-sharing $4d^{5}$ NN ligand cages is RuCl$_{3}$ under high pressure \cite{Bastien_2018},
where a strong stabilization of the singlet state is also found for certain Ru$-$Ru bonds. The present findings on face-sharing
octahedra as encountered in BaIrO$_{3}$ and BaRhO$_{3}$ and recent results on RuCl$_{3}$ \cite{Bastien_2018}
only provide additional motivation for even more detailed electronic-structure calculations on both edge- and face-sharing compounds,
with main focus on the subtle interplay among strong spin-orbit interactions, direct $d-d$ orbital overlap and bonding, and couplings to the lattice degrees of freedom.
%The results suggest the critical role of lattice degrees of freedom along with SOC dictate the magnetic interactions of the heavy transition-metal oxides.
%Although the material model employed here is oversimplified,
%we expect that the nearest-neighbor model dominates over the longer range exchanges, and including the latter should not alter the results qualitatively.
%We further emphasize that understanding the idealized model introduced in this work is the first step towards a complete picture of the oxides with face-sharing octahedra.

%the virtue of our results is that it can give physical insight into
%........\\
%........\\

%%%%%%%%%%%%%%%%%%%%%%%%
%%%             Acknowledgements              %%%
%%%%%%%%%%%%%%%%%%%%%%%%
\section{Acknowledgements}
Calculations were performed at the High Performance Computing Center (ZIH) of the Technical
University Dresden (TUD). We thank U.~Nitzsche for technical assistance.
We acknowledge financial support from the German Science Foundation (Deutsche Forschungsgemeinschaft,
DFG --- HO-4427/2 and SFB-1143) and thank  D. I. Khomskii, V. M. Katukuri, N. A. Bogdanov, and S.-L. Drechsler for instructive discussions.

\clearpage
%\newpage\vspace{4mm}
%\appendix

\begin{appendix}

\setcounter{table}{0}
\renewcommand{\thetable}{A\arabic{table}}
\setcounter{figure}{0}
\setcounter{subfigure}{0}
\renewcommand{\thefigure}{A\arabic{figure}}
\setcounter{equation}{0}
\renewcommand{\theequation}{A\arabic{equation}}

\section{Appendix A: Effective spin model}\label{Appendix_A}

To put in perspective the general trends obtained in the QC calculations for the isotropic exchange coupling $J$,
an effective two-site model is analyzed here.
Two different mechanisms are considered:
($a$) direct $t_{2g}-t_{2g}$ hopping $\sim$$t^{dd}$ and ($b$) indirect processes via the bridging oxygens, $\sim$$t^{dpd}$.
The three bridging oxygens within the median $xy$ mirror plane are denoted as O$_{n}$, with $n\in\{1,2,3\}$.
For each metal ion at sites $l=A,B$, the trigonal CF term ${\cal H}_{\textrm{CF},t}$ splits the $t_{2g}$ orbital states into a two- and an one-dimensional subspace with basis states $|e_{1l}\rangle$, $|e_{2l}\rangle$  and  $|a_{l}\rangle$, respectively \cite{book_S_Sugano}. In what follows, these hole states are denoted as $|\phi_{\mu l}\rangle$, with $\mu=1,2$ for  $|e_{1,2}\rangle$, $\mu=3$ for $|a\rangle$, and
the creation (annihilation) operator $\phi^{\dagger}_{\mu l, \sigma}$ ($\phi_{\mu l, \sigma}$), 
where the spin variable $\sigma =\pm 1/2$ is added. 
Restricted to this low-energy orbital space, a single hole is described by the effective orbital angular momentum operator $L=1$ \cite{book_S_Sugano} in the spin-orbit term ${\cal H}_{\textrm{SO}}=\lambda \sum_l {\bf L}_l {\bf S}_l$; here, ${\bf S}$ is the spin-1/2 operator and $\lambda >$0. Altogether, relevant intra-atomic interactions are collected in the effective Hamiltonian
${\cal H}_{\textrm{at}} = {\cal H}_{\textrm{CF},t} + {\cal H}_{\textrm{SO}} + {\cal H}_{U}$, where
\begin{eqnarray}
{\cal H}_{\textrm{CF},t} &+& {\cal H}_{\textrm{SO}} = \frac{\Delta_t}{3}\sum\limits_{l}\sum\limits_{\mu, \sigma}(\delta_{\mu1}+\delta_{\mu2}-2\delta_{\mu3})\phi^{\dagger}_{\mu l, \sigma}\phi_{\mu l, \sigma}  \nonumber \\
&&+\lambda\sum\limits_{\mu \mu'}\sum\limits_{\sigma \sigma'}[{\bf L}_l]_{\mu \mu'}\cdot [{\bf S}_l]_{\sigma \sigma'}\phi^{\dagger}_{\mu l, \sigma}\phi_{\mu' l, \sigma'}. \label{a1}
\end{eqnarray}
Here, $[{\bf L}_l]_{\mu \mu'}=\langle\phi_{\mu l} |{\bf L}_l| \phi_{\mu' l}\rangle$ while $\Delta_t$ is the trigonal CF splitting between $|e_{1,2}\rangle$ and $|a\rangle$ {\lq\! local \!\rq} states. As stated in the main text, without SOC the calculated hole ground state of Ir$^{4+}$/Rh$^{4+}$ ions is of $e$-orbital character, which 
means $\Delta_t<0$. The term $ {\cal H}_{U}$ includes the leading on-site Coulomb interaction
\begin{equation}
 {\cal H}_{U}=\frac{U}{2}\sum\limits_{l}\sum\limits_{\mu \mu'}\sum\limits_{\sigma \sigma'}n_{\mu l, \sigma}n_{\mu' l, \sigma'},     \label{a2}
\end{equation}
where $n_{\mu l, \sigma}=\phi^{\dagger}_{\mu l, \sigma}\phi_{\mu l, \sigma}$. The approximation of assuming the Coulomb $U$ in the above expression to be independent of the orbital indices $\mu$ and $\mu'$ simplifies the calculation of the isotropic exchange $J$ but excludes obtaining an estimate for the weaker anisotropic exchange $\Gamma$.

Within the one-hole sector and in the cubic limit $\Delta_{t}\to0$,  ${\cal H}_{\textrm{at}}$ is reduced to ${\cal H}_{\textrm{SO}}$. As well known  \cite{book_S_Sugano}, the {\lq original\rq} six $L=1$ atomic states $|L_z=0,\pm 1; \sigma=\pm 1/2\rangle$ are split by ${\cal H}_{\textrm{SO}}$ into the Kramers doublet $|j=1/2; m=\pm1/2\rangle$ and the quartet $|j=3/2; m=\pm1/2,\pm 3/2\rangle$, whose eigenvalues are $-\lambda$ and $\lambda/2$, respectively; here, the site index $l$ is omitted for brevity. When lowering the CF symmetry to trigonal, i.e., $\Delta_t \not=0$, states 
with the same $m(=\pm1/2)$,
$|j=1/2; m\rangle$ and $|j=3/2; m\rangle$, are admixed. By solving the corresponding  2$\times$2 problem, 
the resulting doublet wave-functions are $|\psi_1(m=\pm 1/2)\rangle=c_1|1/2; m\rangle \pm c_2|3/2; m\rangle$ and $|\psi_2 (m=\pm 1/2)\rangle=\mp c_2|1/2; m\rangle + c_1|3/2; m\rangle$, where $c_{1,2}=\left[1/2(1\pm A/\sqrt{A^2+B^2})\right]^{1/2}$ and $A=3-\delta$, $B=2\sqrt{2}\delta$, $\delta=2\Delta_t/3\lambda$. The corresponding eigenvalues are $E_{1,2}=(-\lambda/4)\left[1+3\delta\pm3\sqrt{1-2\delta/3+\delta^2} \right]$ and, since  $\lambda >0$, $E_1<E_2$. The energy of the remaining doublet, $|\psi_3(m)\rangle= |j=3/2; m=\pm 3/2\rangle$, is $E_3=\lambda/2$.

In general, the initial and new basis states are related by an unitary transformation with the rotation matrix $U_{km,\mu\sigma}$ (here, the site index $l$ is restored):
\vspace*{-3mm}
\begin{equation}
 |\phi_{\mu l};\sigma\rangle=\sum\limits_{k,m}|\psi_{kl}(m)\rangle U_{km,\mu\sigma}. \label{a3}\vspace*{-2mm}
\end{equation}
Close inspection of the above expressions for the $E_k$ energy levels ($k=1,2,3$) shows that the ground-state doublet ($k=1$) is well separated from the excited ones ($k=2,3$) for any $\Delta_t$; the low-energy magnetic properties of the system are therefore described by pseudospin-1/2 states $|\psi_{1l}(m)\rangle$. Projection on the low-energy subspace consists in retaining in Eq.(\ref{a3}) the term $k=1$ only, which reads with the replacement $m\to s$ as
\begin{eqnarray}
 &&|a_l;\sigma=\pm 1/2\rangle\to\mp \cos\gamma \ |\psi_{1l}(s=\pm 1/2)\rangle,\nonumber \\
 &&|e_{1l};\sigma=\pm 1/2\rangle\to \frac{1}{\sqrt{2}}\sin\gamma \ |\psi_{1l}(s=\pm 1/2)\rangle,\label{a4} \\
 &&|e_{2l};\sigma=\pm 1/2\rangle\to \pm \frac{i}{\sqrt{2}}\sin\gamma \ |\psi_{1l}(s=\mp 1/2)\rangle,\nonumber
\end{eqnarray}
where $\cos\gamma = (c_1-\sqrt{2}c_2)/\sqrt{3}$ and $\sin\gamma = (\sqrt{2}c_1+c_2)/\sqrt{3}$. 
In the following, the creation (annihilation) of state $ |\psi_{1l}(s)\rangle$ is associated with the operator $ \psi^\dagger_{1l, s}$ ($ \psi_{1l, s}$).
Projected onto the pseudospin-1/2 subspace, the Coulomb interaction ${\cal H}_U$ takes the Hubbard-like form ${\cal H}_U=U\sum_l n_{1l, \uparrow}n_{1l, \downarrow}$. Actually, the unitary transformation (\ref{a3}) yields $\sum_{\mu,\sigma }n_{\mu l, \sigma}=\sum_{k,m}n_{k l, m}$, where only the term $k=1$ is kept. 
%Below, the model is complemented with hopping terms projected onto the low-energy subspace with the use of Eqs.(\ref{a4}).

In case of face-sharing octahedra, the relatively short $\textrm{M}_A-\textrm{M}_B$ distance dictates inclusion of the direct $t_{2g}-t_{2g}$ hopping term
\begin{equation}
 {\cal H}_{hop}^{dd}=\sum\limits_{\mu, \sigma}t^{dd}_{\mu\mu}(\phi^{\dagger}_{\mu A, \sigma}\phi_{\mu B, \sigma} + H.c.). \label{a5}
\end{equation}
The precise structure of $ {\cal H}_{hop}^{dd}$ is determined by symmetry arguments that require that ($a$) the off-diagonal hopping is zero, i.e., $t^{dd}_{\mu\mu'}=0$ if $\mu \not = \mu'$ and ($b$) there are two independent hopping integrals, namely, $t^{dd}_{11}=t^{dd}_{22}\equiv t^{dd}_{e}$ and $t^{dd}_{33}\equiv t^{dd}_{a}$. Projection onto the low-energy subspace then leads to
\begin{equation}
{\cal H}_{hop}^{dd}\simeq  t^{dd}\sum\limits_{s}(\psi^{\dagger}_{1 A, s}\psi_{1 B, s} + H.c.),  \label{a6}
\end{equation}
where $t^{dd}=t^{dd}_{e}\sin^2\gamma  + t^{dd}_{a}\cos^2\gamma $.  Obviously, variation of the M$_A$$-$M$_B$ distance  $d_{\text{MM}}$
gives rise to strong variation of the hopping integral $t^{dd}$.
The $a$-channel contribution $\sim$$t^{dd}_{a}$ is expected to be most sensitive to varying $d_{\text{MM}}$. For instance, according to \cite{book_WA_Harrison}
$ t^{dd}_{a} \sim d_{\textrm{MM}}^{-5}$  .

The treatment of indirect hopping processes via the bridging oxygens is a challenging problem.
The M-O$_3$-M unit should be viewed as a complex molecular-like structure, where superexchange couplings must be analyzed in terms of symmetry-adapted 
molecular orbitals of the O$_{3}$ bridging group. A detailed analysis shows that in the low-energy subspace the indirect hopping term $ {\cal H}_{hop}^{dpd}$ has 
the  same structure as  $ {\cal H}_{hop}^{dd}$, Eq.(\ref{a6}), with the replacement $ t^{dd}\to t^{dpd} = t^{dpd}_{e}\sin^2\gamma  + t^{dpd}_{a}\cos^2\gamma $. The 
hopping integrals $t^{dpd}_{e,a}$ due to second-order processes that occur through  intermediate ligand-hole states
in the $e$- and $a$-channel, respectively,
can be expressed in factorized form as $t^{dpd}_{e,a}\approx [(t^{dp})^2/\Delta_{\textrm{CT}}]F_{e,a}(\theta)$. Here,  $t^{dp}$ and $\Delta_{\textrm{CT}}$ define the characteristic $p-d$ hopping and charge-transfer energy scales. While the factor $F_{e,a}(\theta)$ is strongly dependent on
the angle $\theta$,
the parameter $t^{dp}$ is most sensitive to the metal-oxygen distance $d_{\textrm{MO}}$. According to \cite{book_WA_Harrison}, $t^{dp}\sim d_{\textrm{MO}}^{-7/2}$. Transitions of first-order ($\sim$$t^{dd}_{a,e}$) and second-order type ($\sim$$t^{dpd}_{a,e}$) contribute in each sector  
independently to give the total transfer integral $t= t_{e}\sin^2\gamma + t_a\cos^2\gamma$, where $t_{e,a}=t_{e,a}^{dd} + t_{e,a}^{dpd}$. The resultant hopping Hamiltonian $
{\cal H}_{hop}$ takes the same form as in Eq.(\ref{a6}), but with the replacement $t^{dd}\to t$. The weight factors of the $e$- and $a$- channels are $w_e=\sin^2\gamma$ and $w_a=\cos^2\gamma$, respectively. As discussed above, these factors are controlled by the ratio $\Delta_t/\lambda$.  %Actual values of this parameter obtained in our QC calculations suggest $\tan^2\gamma >$4 and  hence,  the e-channel  prevails partly due to the fact that without SOC the hole %ground state is of the e-orbital character.

It is seen that the generic Hamiltonian ${\cal H}={\cal H}_{at} + {\cal H}_{hop}$ derived above takes the form of an 
effective {\lq single-orbital\rq} Hubbard model operating in the pseudospin-1/2 subspace of NN metal ions.
It can be  treated perturbatively  in the strong correlation regime $t/U$$\ll$$1$,  meaning that excited polar states with two holes on the same metal ion are well 
separated from the low-energy magnetic excitations. In this regime, one immediately obtains as second-order estimate for  the isotropic exchange  $J = 4t^2/U$.
%The validity of this model can be established by comparing its predictions for
%$J$ (see main text and Table \ref{t_eff_mod}) with the $J$ values obtained from spin-orbit 
%QC calculations when changing the local geometry.
%A few more remarks should be added in relation to one complementary AFM and one competing FM contribution to the total $J$. First, there
%are superexchange pathways that include intermediate virtual states with two holes on a bridging oxygen. Their common contribution $\Delta J$  to the total $J$ is of 
%the same sign, $\Delta J > 0$ (AFM),  as that of the leading term  $J=4t^2/U$, but certainly smaller in magnitude.
%A supplementary FM potential exchange,  $\Delta J' < 0$, estimated in the QC approach at  the rAS level is found to be small, as expected. Finally, we also mention
%that the strong correlation regime $U\gg t$ tends to be violated with increasing  $|t|$,  in which case higher-order corrections to $J$ should be taken into account.\\
%

\section{Appendix B: Intersite magnetic couplings}\label{Appendix B}
%{\bf \centerline{Appendix A: Intersite magnetic coupling}}
%The NN magnetic coupling constants were obtained on the basis of multireference configuration-interaction (MRCI) calculations \cite{book_QC_00} on units
%of two face-sharing IrO$_{6}$ (RhO$_{6}$) octahedra. We also included the closest 3 Ba ions around the sharing O$_{3}$ facet in the cluster
%since it is important to accurately describe the charge distribution at sites in the immediate neighborhood \cite{deGraaf_1998,Liviu_2011,Liviu_2007}.
%In order to keep the charge neutrality, we incorporated two extra Ba ions which are adjacent to TM ions along the $z$ axis.
%
All computations were performed with the {\sc {molpro}} quantum chemistry package \cite{Molpro12}. Energy-consistent relativistic pseudopotentials were used for 
the Ir \cite{basis_set_Os} and Rh \cite{basis_setY-Pd} ions. For the Ir/Rh sites, the valence orbitals were described by basis sets of tripe-zeta
quality supplemented with two $f$ polarization functions \cite{basis_set_Os,basis_setY-Pd}. For the ligand O's bridging the two magnetically active Ir (Rh) ions, 
quintuple-zeta valence basis sets and four $d$ polarization functions were applied \cite{basis_set_O}. The other O's were modeled by triple-zeta
valence basis sets \cite{basis_set_O}. The five Ba ions were  modeled by Ba$^{2+}$ {\textquoteleft total-ion\textquoteright}  pseudopotentials (TIP's) supplemented 
with a single $s$ function \cite{basis_set_Ba}.
We used interatomic distances as derived by E. Stitzer {\it{et al.}} \cite{Stitzer_BaRhO3_2004}.

The mapping of the \textit{ab initio} quantum chemistry data onto the effective spin model defined by \,(\ref{eq_spin_H}) implies the
lowest four spin-orbit states associated with the different possible couplings of two NN pseudospins-$1/2$. 
%To derive numerical values for
%all effective spin interactions allowed by symmetry in Eq.\,(\ref{eq_spin_H}), we additionally consider the Zeeman coupling:
In order to safely identify the singlet and triplet components \cite{Ir214_bogdanov_15}, we also consider the Zeeman coupling
%
%\begin{equation}
%\label{ H_Zeeman}
%{\cal \hat{H}}^{Z}_{i,j} =   \sum_{q=i,j} \mu_{\!B} ( \textbf{L}_{q} + \textit{g}_{e} \textbf{S}_{q}) \cdot \textbf{H}
%\end{equation}
%
\begin{equation}
\label{ H_Zeeman}
{\cal \hat{H}}^{Z}_{} =   \sum_{l=A,B} \mu_{\!B} ( \textbf{L}_{l} + \textit{g}_{e} \textbf{S}_{l}) \cdot \textbf{H},
\end{equation}
where $\textbf{L}_{l}$ and $\textbf{S}_{l}$ are angular-momentum and spin operators at a given Ir/Rh site, while $\textit{g}_{e}$ and
${\mu}_{\!B}$ stand for the free-electron Land$\acute{\text{e}}$ factor and Bohr magneton, respectively. Each of the resulting matrix element
computed at the quantum chemistry level is assimilated to the corresponding matrix element of the effective spin Hamiltonian. This
one-to-one correspondence between $\textit{ab initio}$ and effective-model matrix elements enables a clear assignment
of each magnetically active spin-orbit CASSCF/MRCI state and determination of all couplings constants \cite{Ir214_bogdanov_15}.
Effective coupling constants at the rAS+SOC, CAS+SOC, and CI+SOC levels are listed in Table\;\ref{App_Ir_angl}, Table\;\ref{App_Ir_dist},
Table\;\ref{App_Rh_angl}, and Table\;\ref{App_Rh_dist}, complementary to tables and figures in the main text.

\begin{table}[!t]
\caption{
Energy splittings for the lowest four spin-orbit states of two face-sharing NN IrO$_{6}$ octahedra and the corresponding effective
coupling constants obtained from rAS+SOC, CAS+SOC, and MRCI+SOC calculations. 
For each geometry the Ir-Ir distance was fixed to 2.63{\AA}.
For $\theta$=90$^{\circ}$, for instance, the $J$ values without SOC by rAS, CAS, and CI are 0.04, 64.7, and 116.4 meV, respectively.
Distances between the Ir sites and the bridging O's, $d$(Ir-O), are provided within brackets.
The data are presented as complementary to those in Table\;\ref{Ir_coupling_angl} (all values in meV).
}
\hspace{-5mm}
\begin{tabular}{cccc}
\hline
\hline\\[-0.30cm]
 \hspace{0mm} $\theta$, $d$(Ir-O) \hspace{4mm}  &\hspace{0mm} rAS+SOC \hspace{0mm} &\hspace{0mm} CASSCF+SOC  \hspace{0mm}  &MRCI+SOC\\%%($\times\!2$)\\
\hline
\\[-0.20cm]
     $70.5^{\circ}$ (2.27\AA):       &                   &          &      \\
         $\Psi_{\!S}$                & 9.9               &  0.0        & 0.0  \\
         $\Psi_{\!2}$                & 0.0               & 24.8       & 58.2\\
         $\Psi_{\!1}$                & 0.5               & 34.9       & 73.4  \\
         $\Psi_{\!3}$                & 0.5               & 34.9       & 73.4  \\
         ${\bar J},{\bar \Gamma} $          & $-$9.9, 1.1      & 24.8, 20.3  & 58.2, 30.3 \\[0.3cm]
     $75^{\circ}$ (2.16\AA):         &                   &           & \\
          $\Psi_{\!S}$               & 12.8           &  0.0          & 0.0\\
          $\Psi_{\!2}$               & 0.0               & 36.3       & 73.9\\
          $\Psi_{\!1}$               & 0.0               & 41.6       & 81.8\\
          $\Psi_{\!3}$               & 0.0               & 41.6       & 81.8 \\
          ${\bar J},{\bar \Gamma} $         & $-$12.8, 0.0   &36.3, 10.5   & 73.9, 15.9   \\[0.3cm]
      $80^{\circ}$ (2.04\AA):         &                   &           & \\
	$\Psi_{\!S}$         &15.2         & 0.0        &   0.0  \\
	$\Psi_{\!2}$         & 0.4         &72.0        & 123.3  \\
	$\Psi_{\!1}$         & 0.0         &74.0        & 126.5  \\
	$\Psi_{\!3}$         & 0.0         &74.0        & 126.5  \\
	${\bar J},{\bar \Gamma}$              &$-$14.9, $-$0.7    &72.0, 4.1      &123.3, 6.3  \\[0.3cm]
     $85^{\circ}$ (1.94\AA):         &                   &            &\\
          $\Psi_{\!S}$               & 16.2           &  0.0          & 0.0\\
          $\Psi_{\!2}$               & 0.3               &149.4      & 225.8\\
          $\Psi_{\!1}$               & 0.0               &150.1      & 226.9\\
          $\Psi_{\!3}$               & 0.0               &150.1      & 226.9\\
          ${\bar J},{\bar \Gamma} $         & $-$15.9, $-$0.6   &149.4, 1.3   &225.8, 2.1  \\[0.3cm]
     $90^{\circ}$ (1.86\AA):         &                   &            &\\
          $\Psi_{\!S}$               & 17.0              &  0.0        & 0.0\\
          $\Psi_{\!2}$               & 0.0               &293.6        & 417.5\\
          $\Psi_{\!1}$               & 0.3               &294.2        & 418.5\\
          $\Psi_{\!3}$               & 0.3               &294.2        & 418.5\\
          ${\bar J},{\bar \Gamma} $         & $-$17.0, 0.7      &293.6, 1.2  &417.5, 2.0   \\[0.3cm]
\hline
\hline
\end{tabular}
\label{App_Ir_angl}
\end{table}

\begin{table}[t!]
\caption{
Energy splittings for the lowest four spin-orbit states of two face-sharing NN IrO$_{6}$ octahedra and the corresponding effective
coupling constants obtained from rAS+SOC, CAS+SOC, and MRCI+SOC calculations. The relative distances from the O ligands to the $z$ axis
were fixed to 1.57 \AA{}. The data are presented as complementary  to those in Fig.\;\ref{J_Ir_dist} (all values in meV).
}
\hspace{-5mm}
\begin{tabular}{cccc}
\hline
\hline\\[-0.30cm]
 \hspace{2mm} $d$(Ir-Ir) \hspace{2mm}  &\hspace{2mm} rAS+SOC \hspace{2mm} &\hspace{2mm} CASSCF+SOC  \hspace{2mm} & MRCI+SOC \\%%($\times\!2$)\\
\hline
\\[-0.20cm]
     2.50\AA :                                                  &                   &               &  \\
         $\Psi_{\!S}$                                        & 22.0              &  0.0         &  0.0    \\
         $\Psi_{\!2}$                                        & 0.5               & 188.6     &   329.7  \\
         $\Psi_{\!1}$                                        & 0.0               & 198.1     &   345.0 \\
         $\Psi_{\!3}$                                        & 0.0               & 198.1     &   345.0\\
         ${\bar J},{\bar \Gamma} $          & $-$21.5 $-$0.93      & 188.6, 18.9  &  329.7, 30.5\\[0.3cm]
     2.56\AA :                           &                     &            \\
          $\Psi_{\!S}$                                      & 18.3              &  0.0         & 0.0 \\
          $\Psi_{\!2}$                                      & 0.4                & 106.7       & 187.2 \\
          $\Psi_{\!1}$                                      & 0.0                & 111.2        & 194.3\\
          $\Psi_{\!3}$                                      & 0.0                & 111.2       &  194.3\\
          ${\bar J},{\bar \Gamma} $         & $-$17.9, $-$0.8      & 106.7, 8.9    & 187.2, 14.2  \\[0.3cm]
     2.63\AA :                           &                     &            \\
          $\Psi_{\!S}$                                      & 15.2              &  0.0         & 0.0 \\
          $\Psi_{\!2}$                                      & 0.4                & 72.0       & 123.3 \\
          $\Psi_{\!1}$                                      & 0.0                & 74.0       & 126.5 \\
          $\Psi_{\!3}$                                      & 0.0                & 74.0       & 126.5 \\
          ${\bar J},{\bar \Gamma} $         & $-$14.9, $-$0.7      & 72.0, 4.1    & 123.3, 6.3  \\[0.3cm]
     2.69\AA :                          &                      &            \\
          $\Psi_{\!S}$                                     & 12.6              &   0.0      & 0.0 \\
          $\Psi_{\!2}$                                     & 0.3                &  54.7      & 90.5\\
          $\Psi_{\!1}$                                     & 0.0                &  55.5      & 91.7\\
          $\Psi_{\!3}$                                     & 0.0                &  55.5      & 91.7\\
          ${\bar J},{\bar \Gamma} $         & $-$12.4, $-$0.54   &54.7, 1.7      & 90.5, 2.2   \\[0.3cm]
     2.76\AA :                         &                    &            \\
          $\Psi_{\!S}$                                    & 10.4              &  0.0       &  0.0\\
          $\Psi_{\!2}$                                    & 0.2                & 44.5       & 72.1\\
          $\Psi_{\!1}$                                    & 0.0                & 44.8       & 72.4\\
          $\Psi_{\!3}$                                    & 0.0                & 44.8       & 72.4\\
          ${\bar J},{\bar \Gamma} $         & $-$10.3, $-$0.4      &44.5, 0.6    & 72.1, 0.5  \\[0.3cm]
\hline
\hline
\end{tabular}
\label{App_Ir_dist}
\end{table}

%%%%%%%%%%%%%
%%     TABLE A-3      %%
%%%%%%%%%%%%%
\begin{table}
\caption{
Energy splittings for the lowest four spin-orbit states of two face-sharing NN RhO$_{6}$ octahedra and the corresponding effective
coupling constants obtained from rAS+SOC,  CAS+SOC, and MRCI+SOC calculations. 
For each geometry the Rh-Rh distance was fixed to 2.58 \AA{} .
Distances between the Rh sites and the bridging O's, $d$(Rh-O), are provided within brackets.
The data are presented as complementary to those in Table\;\ref{Rh_coupling_angl} (all values in meV).
}
\hspace{-5mm}
\begin{tabular}{cccc}
\hline
\hline\\[-0.30cm]
 \hspace{2mm} $\theta$, $d$(Rh-O)\hspace{2mm}  &\hspace{2mm} rAS+SOC \hspace{2mm} &\hspace{2mm} CASSCF+SOC  \hspace{2mm} & MRCI+SOC \\%%($\times\!2$)\\
 \hline
\\[-0.20cm]
     $70.5^{\circ}$ (2.23\AA ):                         &                    &                           &         \\
         $\Psi_{\!S}$                                        & 10.4            &  6.5                    & 0.0    \\
         $\Psi_{\!3}$                                        & 0.0               & 1.0                    & 3.7    \\
         $\Psi_{\!1}$                                        & 0.0               & 1.0                    & 3.7    \\
         $\Psi_{\!2}$                                        & 0.6               & 0.0                    & 1.5    \\
         ${\bar J},{\bar \Gamma} $          & $-$9.9 $-$1.1      & $-$6.5, 2.0        &1.5, 4.4    \\[0.3cm]
     $75^{\circ}$ (2.12\AA):                           &                     &                          &          \\
          $\Psi_{\!S}$                                      & 13.3              &  0.0                  & 0.0     \\
          $\Psi_{\!3}$                                      & 0.0                & 0.1                   & 15.0   \\
          $\Psi_{\!1}$                                      & 0.0                & 0.1                   & 15.0   \\
          $\Psi_{\!2}$                                      & 0.8                & 0.4                   & 15.2   \\
          ${\bar J},{\bar \Gamma} $         & $-$12.5, $-$1.7      & 0.4, $-$0.6    & 15.2, $-$0.4 \\[0.3cm]
      $80^{\circ}$ (2.06\AA):                           &                     &                         &          \\
          $\Psi_{\!S}$                                      & 16.0              &  0.0                  & 0.0     \\
          $\Psi_{\!3}$                                      & 0.0                & 25.4                   & 54.0   \\
          $\Psi_{\!1}$                                      & 0.0                & 25.4                   & 54.0   \\
         $\Psi_{\!2}$                                      & 1.1                & 26.5                   & 55.5   \\
          ${\bar J},{\bar \Gamma} $         & $-$14.9, $-$2.3      & 26.5, $-$2.2    & 55.5, $-$2.8 \\[0.3cm]
     $85^{\circ}$ (1.91\AA):                          &                      &                         &            \\
          $\Psi_{\!S}$                                     & 17.7              &   0.0                  & 0.0      \\
          $\Psi_{\!3}$                                     & 0.0                &  82.4                 & 135.7  \\
          $\Psi_{\!1}$                                     & 0.0                &  82.4                 & 135.7   \\
          $\Psi_{\!2}$                                     & 1.3                &  84.0                 & 137.8   \\
          ${\bar J},{\bar \Gamma} $         & $-$16.4, $-$2.6   &84.0, $-$3.2      & 137.8, $-$4.3 \\[0.3cm]
     $90^{\circ}$ (1.82\AA):                         &                      &                            &             \\
          $\Psi_{\!S}$                                    & 16.5               &  0.0                    &  0.0      \\
          $\Psi_{\!3}$                                    & 0.0                 &165.4                  &  173.4   \\
          $\Psi_{\!1}$                                    & 0.0                 &165.4                  & 173.4   \\
          $\Psi_{\!2}$                                    & 0.9                 &190.5                  & 164.0   \\
          ${\bar J},{\bar \Gamma} $        & $-$15.6, $-$1.8    &190.5, $-$50.2   &  164.0, 18.8 \\[0.3cm]
\hline
\hline
\end{tabular}
\label{App_Rh_angl}
\end{table}

%%%%%%%%%%%%%
%%     TABLE A-4      %%
%%%%%%%%%%%%%
\begin{table}
\caption{
Energy splittings for the lowest four spin-orbit states of two face-sharing NN RhO$_{6}$ octahedra and the corresponding effective
coupling constants obtained from rAS+SOC,  CAS+SOC, and MRCI+SOC calculations.
The relative distances from the O ligands to the $z$ axis are fixed to 1.54 \AA. The data are presented as complementary to those in Fig.\;\ref{J_Rh_dist} (all values in meV).
}
\hspace{-5mm}
\begin{tabular}{cccc}
\hline
\hline\\[-0.30cm]
 \hspace{2mm} $d$(Rh-Rh)\hspace{2mm}  &\hspace{2mm} rAS+SOC \hspace{2mm} &\hspace{2mm} CASSCF+SOC  \hspace{2mm} & MRCI+SOC \\%%($\times\!2$)\\
 \hline
\\[-0.20cm]
      2.45\AA :                         &                    &                       &         \\
         $\Psi_{\!S}$                                        & 22.0               & 0.0                     & 0.0    \\
         $\Psi_{\!3}$                                        & 0.0                 & 49.5                   & 157.6    \\
         $\Psi_{\!1}$                                        & 0.0                 & 49.5                   & 157.6    \\
         $\Psi_{\!2}$                                        & 1.6                 & 50.0                   & 158.1    \\
         ${\bar J},{\bar \Gamma} $          & $-$20.4 $-$3.2      & 50.0, $-$1.1       &158.1, $-$1.1    \\[0.3cm]
     2.51\AA :                           &                     &                          &          \\
          $\Psi_{\!S}$                                      & 18.8               &  0.0                   & 0.0     \\
          $\Psi_{\!3}$                                      & 0.0                 & 30.1                   & 73.2   \\
          $\Psi_{\!1}$                                      & 0.0                 & 30.1                   & 73.2   \\
          $\Psi_{\!2}$                                      & 1.4                 & 31.1                   & 74.5   \\
          ${\bar J},{\bar \Gamma} $         & $-$17.5, $-$2.7    & 31.1, $-$2.1    & 74.5, $-$2.5 \\[0.3cm]
      2.58\AA :                           &                     &                         &          \\
          $\Psi_{\!S}$                                      & 16.0              &  0.0                  & 0.0     \\
          $\Psi_{\!3}$                                      & 0.0                & 25.4                  & 54.0   \\
          $\Psi_{\!1}$                                      & 0.0                & 25.4                  & 54.0   \\
          $\Psi_{\!2}$                                      & 1.1                & 26.5                  & 55.5   \\
          ${\bar J},{\bar \Gamma} $         & $-$14.9, $-$2.3   & 26.5, $-$2.2     & 55.5, $-$2.8 \\[0.3cm]
     2.64\AA :                          &                      &                         &            \\
          $\Psi_{\!S}$                                     & 13.6              &   0.0                  & 0.0      \\
          $\Psi_{\!3}$                                     & 0.0                &  23.8                 & 47.0 \\
          $\Psi_{\!1}$                                     & 0.0                &  23.8                 & 47.0   \\
          $\Psi_{\!2}$                                     & 1.0                &  24.7                 & 48.3  \\
          ${\bar J},{\bar \Gamma} $         & $-$12.6, $-$1.9   &24.7, $-$2.0      & 48.3, $-$2.6 \\[0.3cm]
     2.71\AA :                         &                      &                            &             \\
          $\Psi_{\!S}$                                    & 11.5               &  0.0                   &  0.0 \\
          $\Psi_{\!3}$                                    & 0.0                 & 22.6                  &  42.6 \\
          $\Psi_{\!1}$                                    & 0.0                 & 22.6                  &  42.6  \\
          $\Psi_{\!2}$                                    & 0.8                 & 23.5                  &  43.7  \\
          ${\bar J},{\bar \Gamma} $        & $-$10.7, $-$1.6    & 23.5, $-$1.7     &  43.7, $-$2.2 \\[0.3cm]
\hline
\hline
\end{tabular}
\label{App_Rh_dist}
\end{table}

\clearpage

%%%%%%%%%%%%%%
%%     FIGURE A-3.    %%
%%%%%%%%%%%%%%
%%%\begin{figure}
%%%\centering
 %%%\hspace{-0.5cm}
 %%%\subfigure{ \hspace{-0.5cm}
%%%\includegraphics[width=0.3\textwidth,natwidth=20,angle=-90]{cas_J_Ir.pdf}
%%%\label{splitting_dist}}
 %%%\caption{Dependence on TM-O-TM bond angles of the CASSCF $J$ without SOC for face-sharing octahedra Ir$_{2}$O$_{9}$ and Rh$_{2}$O$_{9}$ fragments.
%%%}
%%%\label{cas_J}
%%%\end{figure}
%
%%%%%%%%%%%%%
%%      TABLE A-3   %%
%%%%%%%%%%%%%
%%%\begin{table}
%%%\caption{
%%%The CASSCF $J$ without SOC for face-sharing octahedra Ir$_{2}$O$_{9}$ and Rh$_{2}$O$_{9}$ fragments at different TM-O-TM bond angles (all in meV).
%%%}
%%%\hspace{-5mm}
%%%\begin{tabular}{ccc}
%%%\hline
%%%\hline\\[-0.30cm]
 %%%\hspace{0mm} $\theta$/$\theta^{\prime}$  \hspace{4mm}   &\hspace{0mm} Ir$_{2}$O$_{9}$  \hspace{0mm}  &  Rh$_{2}$O$_{9}$ \\%%($\times\!2$)\\
%%%\hline
%%%\\[-0.20cm]
%%%     $70.5^{\circ}$      &     0.5              &     $-$2.3         \\  [0.15cm]
%%%      $75^{\circ}$        &     $-$2.5            &     2.8       \\ [0.15cm]
%%%     $80^{\circ}$        &   27.4                &     30.7      \\ [0.15cm]
%%%      $85^{\circ}$       &  49.9                 &     40.4     \\   [0.15cm]
%%%     $90^{\circ}$        &    64.7               &      54.9    \\  [0.15cm]
%%%  \hline
%%% \hline
%%%\end{tabular}
%%%\label{App_Ir_angl}
%%%\end{table}

\end{appendix}

\clearpage
%\bibliography{biblio_Mar23}

\end{document}